\documentclass[
preprint,titlepage,showpacs,preprintnumbers,amsmath,amssymb,noshowkeys,superscriptaddress]{revtex4}
\usepackage{graphicx}
\usepackage{rotating}
\usepackage{epsfig}
\usepackage{color}
\usepackage{array}
\usepackage[dvipsnames]{xcolor}
\usepackage[utf8]{inputenc}
\usepackage[T2A]{fontenc}

\begin{document}
	
\title{Study of Nucleon Charge-Exchange Processes at $^{12}$C \\ Fragmentation
with an Energy of 300~MeV/Nucleon}
\author{\firstname{A.A.}~\surname{Kulikovskaya}}
\email[E-mail: ]{annkull316@mail.ru}
\affiliation{NRC «Kurchatov Institute», Moscow, 123182, Russia}%
\author{\firstname{M.A.}~\surname{Martemianov}}
\affiliation{NRC «Kurchatov Institute», Moscow, 123182, Russia}

\begin{abstract}
The search of reactions with nucleon charge-exchange was performed on the FRAGM fragment-separator of the TWAC accelerator complex at fragmentation of carbon nuclei with an energy of 300~MeV/nucleon on a thin beryllium target. The experimental setup, located at an angle of 3.5$^\circ$ to the incident beam, had high momentum resolution. Differential cross sections were measured for $^{11}$Be, $^{12}$B and $^{12}$Be as function of the nuclei momentum. The experimental data were compared with theoretical predictions of various models of nucleus-nucleus interactions and other experimental results. Measurements of nucleon charge exchange processes in this energy region was carried out for the first time. New results were obtained to test theoretical models of nucleus-nucleus interactions.
\end{abstract} 

		
\maketitle

\section{INTRODUCTION}
	
Investigation of underlying mechanisms of nucleus–nucleus interactions is one of the important areas of research in modern experimental physics. Considerable attention is paid to the issues of a phenomenologically accurate description of the processes of nuclear fragmentation, including the mechanisms of their producing and the features of the nucleus structure, necessary in such fields of application as heavy ion therapy, calculations of radiation protection and the producing of radioactive ion 
beams~\cite{PaperIntr1}.

Fragmentation processes of nucleon charge exchange were rather weak studied experimentally, despite the fact that first works appeared in the 80s of last century~\cite{PaperIntr2}. From the experimental point of view, this is due to the difficulties of identifying these processes in the region of heavy nuclei and the small cross section values in the region of light nuclei. As a result of such reactions, the number of protons or neutrons in the nascent fragment increases compared to the projectile nucleus. The theoretical description of charge-exchange processes is usually based on the mechanism of meson interaction between the nucleons of projectile and target nuclei, namely: at low and intermediate energies of the projectile nucleus, the dominant process is the effect of quasi-elastic scattering. With increase of the energy of projectile nucleus, the interaction of nucleons begins to be accompanied by the production of resonance states. From a practical point of view, charge-exchange reactions serve as a tool for studying a fairly wide range of physical problems. The corresponding experimental data are important for assessing the role of meson exchanges, nucleon-nucleon correlations, modification of baryon resonances in the nuclear medium,
spin-isospin nuclear excitations, and are also necessary for calculating the nuclear matrix elements of neutrinoless double beta decay~\cite{PaperIntr3}. The mechanism of charge exchange between nucleons is a promising tool for nuclear fusion far from stability and hypernuclei.

The creation of large heavy-ion accelerator complexes such as GSI (Germany) and RIKEN (Japan) allowed to perform an 
investigation of charge-exchange processes for the region of medium-mass and heavy nuclei~\cite{PaperIntr4,PaperIntr5}. For example, at the FRS facility in GSI, during the fragmentation of $^{112}$Sn and $^{208}$Pb as a result of the charge exchange reaction, the isotopes $^{112}$Ir and $^{208}$Bi were observed respectively at an energy of 1~GeV/nucleon~\cite{PaperIntr6}. 
The cross section of these fragments turned out to be quite large and comparable with the cross sections for fragmentation proceeding without charge exchange. Analysis of the missing-energy spectra demonstrated the ratios between contributions from quasi-elastic and inelastic interactions of ions in the charge exchange process. Moreover the collision energy in the inelastic interaction was enough to generate an isobaric state. The theoretical analysis of the data made it possible to obtain good agreement with the experiment for the first time. There is a large number of works devoted to the theoretical aspects of both single and double nucleon charge exchange in processes of nuclear fragmentation~\cite{PaperIntr3,PaperIntr7,PaperIntr8,PaperIntr9}.

One of the advantages of the FRAGM experiment was its high momentum resolution, which provided high-precision data on ion fragmentation in the energy range available at the TWAC accelerator complex~\cite{PaperIntr10}. In this experiment, a large amount of data on carbon ion fragmentation was collected on the targets from beryllium to tantalum and in a wide range of projectile nucleus kinetic energies from 0.3 to 3.2~GeV/nucleon, as well as in a wide energy range of the produced fragments from hydrogen isotopes to isotopes of the projectile carbon nucleus. A special feature of the FRAGM experiment is that it measures the momentum spectra of all long-lived fragments of beryllium and boron produced with and without nucleon charge exchange. The purpose of this review is to present the experimental results to search the isotopes formed in nucleon charge-exchange reactions.
The reaction studied was \mbox{$^9$Be~($^{12}$С,~$f$\hspace{0.2em})~X}, where $f$ -- required fragments: $^{11}$Ве, $^{12}$В, $^{12}$Ве and $^{12}$N.The differential cross sections for the yields of obtained fragments were measured as a function of the fragment momentum. The obtained data were compared with various models of nucleus-nucleus interactions, and also were studied the shapes of fragment momentum spectra in the rest frame of the projectile nucleus and then their consistency to other experimental results and predictions of statistical models. There are only a few similar experiments performed in the energy range of 1--2~GeV/nucleon and only for isobaric transitions. Similar measurements at an energy of 300~MeV/nucleon were measured for the first time.	
	
\section{REVIEW OF REACTIONS WITH NUCLEON CHARGE-EXCHANGE}

Charge-exchange reactions include processes in atomic nuclei, when only the charge of nucleus changes, but the total number of nucleons is preserved. Also, there is another representation of such processes, when the number of neutrons or protons in the nucleus increases by one, which can be called a reaction that occurs with a single charge exchange of nucleons. In the history of studying these reactions, the discovery of a special process, called beta decay, can be considered as the beginning, at a time when modern concepts of the atom structure was not yet formed. The creation and development of accelerator technology gave impulse to a more detailed study of nuclear charge-exchange reactions. Initially, studies of charge-exchange processes were carried out on accelerated proton beams. ($p$, $n$) reactions were studied, which made a significant contribution to the study of excited states, isobaric analog states, reaction mechanisms and nuclear structure, and to the calculation of nuclear charge radii. 

A significant difficulty at this stage was measuring the energy spectra of neutrons at high energies. Therefore, the quasi-elastic process ($^3$He, $t$) is under increased interest, since registration of tritons is a simpler task. 
Investigation such process at low and intermediate energies provides important information about the interaction of nucleons in nuclei and the properties of nuclear fields~\cite{PaperReview1,PaperReview2,PaperReview3}. For describing quasi-elastic processes, optical models are often used, in which the particle scattering on a nucleus consisting of many nucleons is described as the passage of an projectile wave through a medium whose properties are determined by the optical potential. The parameters of such potential are selected from the condition of agreement between calculated and experimental data. Analysis of elastic and quasi-elastic scattering on nuclei showed the difference between neutron and proton potentials. Thus, the cross sections of charge-exchange reactions should be determined by the isospin potential and depend on excess of the neutrons. 

Such processes on $^{45}$Sc, $^{194}$Pt and $^{197}$Au nuclei at energies of 10 -- 29 MeV were studied using the $^3$He ion beam of the cyclotron at the Institute of Nuclear Physics (Czech)~\cite{PaperReview4,PaperReview5,PaperReview6}. It was shown, that the population of excited states must depend on the excess of neutrons in the target nucleus. Charge exchange reactions also serve as a method for studying the structure of atomic nucleus. The cross section of such reaction depends on the differences in surface distribution of the nucleon density. Therefore, these charge exchange processes where the nuclei have a structure with a neutron halo are under increased interest.

For intermediate energies, the study of charge-exchange reactions is aimed to investigate a spin-isospin excitations of the target nucleus. The excitation energy spectrum of $^{12}$B was measured at the SMART experimental setup (RIKEN) in the reaction $^{12}$С($^{12}$С,~$^{12}$N)$^{12}$B at an energy of 135 MeV/nucleon~\cite{PaperReview7}. The cross section for the production of the ground state of $^{12}$B was measured. The excitation energy spectra of the $^{12}$N nucleus were obtained at the Indiana 
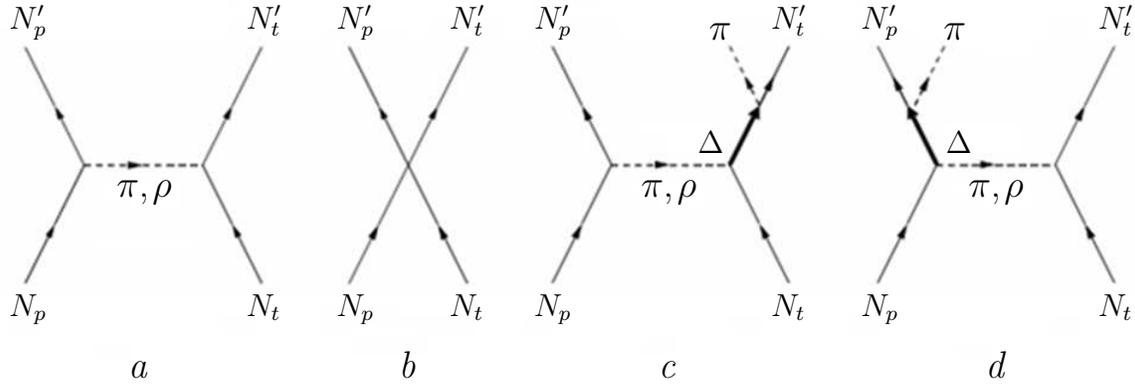
\begin{figure}[!htb]
\includegraphics[scale = 0.45]{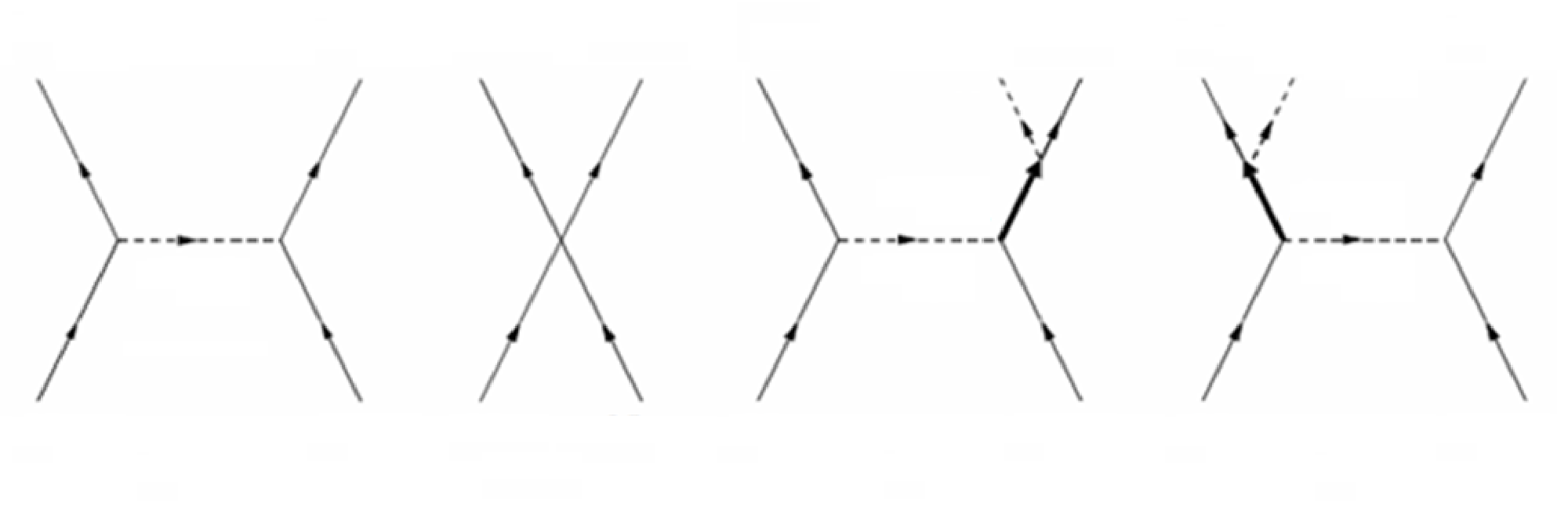}
\setlength{\unitlength}{1.5mm} 
\begin{picture}(400,0)
\put(5,9){\fontsize{12}{12}\selectfont{$N_p$}}
\put(26,9){\fontsize{12}{12}\selectfont{$N_t$}}
\put(34,9){\fontsize{12}{12}\selectfont{$N_p$}}
\put(44,9){\fontsize{12}{12}\selectfont{$N_t$}}
\put(51.5,9){\fontsize{12}{12}\selectfont{$N_p$}}
\put(72.5,9){\fontsize{12}{12}\selectfont{$N_t$}}
\put(80.5,9){\fontsize{12}{12}\selectfont{$N_p$}}
\put(101.5,9){\fontsize{12}{12}\selectfont{$N_t$}}
\put(5,34.5){\fontsize{12}{12}\selectfont{$N_p'$}}
\put(26,34.5){\fontsize{12}{12}\selectfont{$N_t'$}}
\put(34,34.5){\fontsize{12}{12}\selectfont{$N_p'$}}
\put(44,34.5){\fontsize{12}{12}\selectfont{$N_t'$}}
\put(51.5,34.5){\fontsize{12}{12}\selectfont{$N_p'$}}
\put(72.5,34.5){\fontsize{12}{12}\selectfont{$N_t'$}}
\put(80.5,34.5){\fontsize{12}{12}\selectfont{$N_p'$}}
\put(101.5,34.5){\fontsize{12}{12}\selectfont{$N_t'$}}
\put(15.5,3.5){\fontsize{14}{14}\selectfont{{\it a}}}
\put(39.5,3.5){\fontsize{14}{14}\selectfont{{\it b}}}
\put(62.5,3.5){\fontsize{14}{14}\selectfont{{\it c}}}
\put(91.5,3.5){\fontsize{14}{14}\selectfont{{\it d}}}
\put(66.0,23.4){\fontsize{12}{12}\selectfont{$\Delta$}}
\put(88,23.4){\fontsize{12}{12}\selectfont{$\Delta$}}
\put(67,33.5){\fontsize{14}{14}\selectfont{$\pi$}}
\put(87.8,33.5){\fontsize{14}{14}\selectfont{$\pi$}}
\put(14.5,19.6){\fontsize{14}{14}\selectfont{$\pi$,$\hspace{0.1em}\rho$}}
\put(61,19.6){\fontsize{14}{14}\selectfont{$\pi$,$\hspace{0.1em}\rho$}}
\put(90,19.6){\fontsize{14}{14}\selectfont{$\pi$,$\hspace{0.1em}\rho$}}
\end{picture}\vspace{-0.5cm}
\caption{
Schematic view of the processes of quasi-elastic interaction of nucleons of the projectile nucleus ($N_p$) and the target ($N_t$) ({\it a} and {\it b}) and the inelastic channel accompanied by the $\Delta$-isobar production ({\it c} and {\it d})~\cite{PaperIntr6}.} \label{Pict01}
\end{figure}
University (USA) cyclotron in reaction $^{12}$С($p$,~$n$)$^{12}$N~\cite{PaperReview8}. These experimental data are closely related 
to the interpretation of results at the FRAGM setup.

At intermediate and high energies, various baryon resonances are produced as the result of inelastic interaction between the projectile nucleus and the target (Fig.~\ref{Pict01}). In fact, protons and neutrons interact at the periphery of the nucleus via two channels: quasi-elastic and inelastic, which is accompanied by the formation of resonance states. In particular, the inelastic reaction channel, which occurs with the production of $\Delta$-resonance, can be observed at an energy close to 290~MeV. The mechanism of baryon resonance formation inside an excited nucleus still remains a fundamental problem in hadron physics.

The production of $\Delta$-resonance is the most peculiar manifestation of $\pi N$~--~interaction. Since pion exchange plays the main role in nucleon-nucleon interaction, the influence of resonance on the excitation spectrum of nuclear matter is so significant that $\Delta$-isobar can be considered an integral part of nuclear matter along with nucleons~\cite{PaperReview9}.
According to modern theoretical predictions, $\Delta$-isobaric excitation of the nucleus has a collective nature. Due to the short lifetime, resonance quickly decays, and for the nucleon there is a free level in the Fermi distribution for nucleons, and the pion finds another candidate for the production of new isobar. The set of experimental data to investigate $\Delta$-isobars in proton and nuclear charge exchange reactions on various targets shows that the quasi-free mechanism of isobar production is not the main one for these reactions, and the main mechanism of interaction is taking into account as a collective. In accordance with such theoretical assumptions, it was shown that the nuclear energy isobar peak is shifted to lower transferred energies compared to the similar peak in the cross section reaction for a free proton, and the peak width also exceeds its nominal value.
It was also found, that the reason for this shift and the peak spread is associated with the response of nucleus target to isobar production, and not with the type of projectile. Finally, the cross section in the region of inelastic peak in charge-exchange reactions measured on a nuclear target significantly exceeds the similar cross section for the proton target. Important experimental measurements of charge-exchange reactions were carried out on the multipurpose magnetic spectrometer Alpha, located in the beamline of the JINR synchrophasotron at Dubna. A typical charge-exchange reaction ($^3$He, t) in the energy range from 1.45 to 3.6~GeV/nucleon for a proton and carbon target was investigated~\cite{PaperReview2}.

It should be noted, that the modern measurements of $\Delta$-resonance, which were carried out at the FRS fragment-separator and the SIS-18 synchrotron (GSI, Germany)~\cite{PaperIntr4}. The stable nuclei $^{112}$Sn and $^{124}$Sn with an energy of 1~GeV/nucleon were used as projectile nuclei. In the experiment was measured the missing-energy distributions with good accuracy, which made it possible to separate the quasi-elastic and inelastic components of the energy spectrum. For both projectile nuclei, the energy shift for the $\Delta$-isobar peak relative to the same measurements for the hydrogen target was calculated with high precision. The shift was equal to \mbox{63~$\pm$~5 MeV}. An attempt was also made to describe the shape of inelastic peak and to explain this shift using theoretical calculations~\cite{PaperReview10}. It was shown that an isobar, which is generated both in the projectile nucleus and in the target, can contribute to the production of inelastic peak. Peaks from these two mechanisms are shifted in opposite directions, and the resulting shift of the peaks can be determined on the basis of magnitude of $\Delta$-resonance contributions formed in the projectile nucleus and the target.

Double charge exchange reactions of nucleons, when the charge is changed by two units, is still insufficiently studied, especially in the area of reaction mechanisms. There are different theories about whether only one nucleon of the target and the projectile nucleus participates in the reaction or whether interaction of two such nucleons is necessary. Theoretical models describe the mechanism of such interaction in different ways. As already noted, the collision of two nucleons is accompanied by the production of a $\Delta$-resonance, which, interacting with the nucleon, produces a second resonance; the decay of two such isobars is accompanied by a change in the charge of formed fragment by two units~\cite{PaperReview11}. On the other side, nucleon interactions can produce a resonance $N$(1440), which decays via the channel $N$(1440)$\rightarrow$$\Delta\pi{\rightarrow}$$N\pi\pi$ and can explain the double charge exchange of nucleons due to the production of two pions~\cite{PaperReview12}.

\section{EXPERIMENT FRAGM AT THE ACCELERATOR-STORAGE COMPLEX TWAC }
\subsection{\bf Accelerator complex ТWAC. }	
	
The TWAC (TeraWatt ACcumulator) accelerator complex~\cite{PaperSetup1}, created on the basis of the U-10 proton synchrotron, was adapted for ion acceleration and accumulation. In the process its creating the complex, complicated technical and engineering problems were solved~\cite{PaperSetup2}. A new technology was developed for generating high-charged beams of highly charged ions using powerful radiation from a pulse-periodic CO$_2$ laser~\cite{PaperSetup3}. In addition, multiple charge-exchange injection was introduced for accumulating beams of nuclei of various elements. In the process of complex modernization, a new linear ion injector I-3 with energy of 4~MeV, related to a booster synchrotron, was used to accelerate protons up to 10~GeV and ions up to 4~GeV/nucleon, as well as to accumulate nuclei of various elements in the range from 200 to 300~MeV/nucleon~\cite{PaperSetup4}.
The accumulated beam passes through a recharge target once per cycle, so the perturbing effect of the target on a beam is minimal.  	

The accelerator complex allowed to carry out the series of experiments in the field of fundamental heavy-ion physics, high-density energy physics, radiation biology, proton and ion therapy, radiography and other fields. Various adjustments of the complex made it possible to study the mechanisms of ion-ion interactions, the internal structure of the fragmenting nucleus and other aspects. Accelerated beams were used in the following modes: secondary beams obtained during the interaction of accelerated protons or ions with internal targets of the U-10 ring were transported for physical experiments to the experimental hall; proton beams extracted from the U-10 ring in one turn were sent to the medical building for biological research and proton therapy. Acceleration of ions of carbon and further to silver reached relativistic energies at an intensity of  \mbox{10$^7$~--~10$^{10}$s$^{-1}$}.

\subsection{Experimental setup FRAGM. }

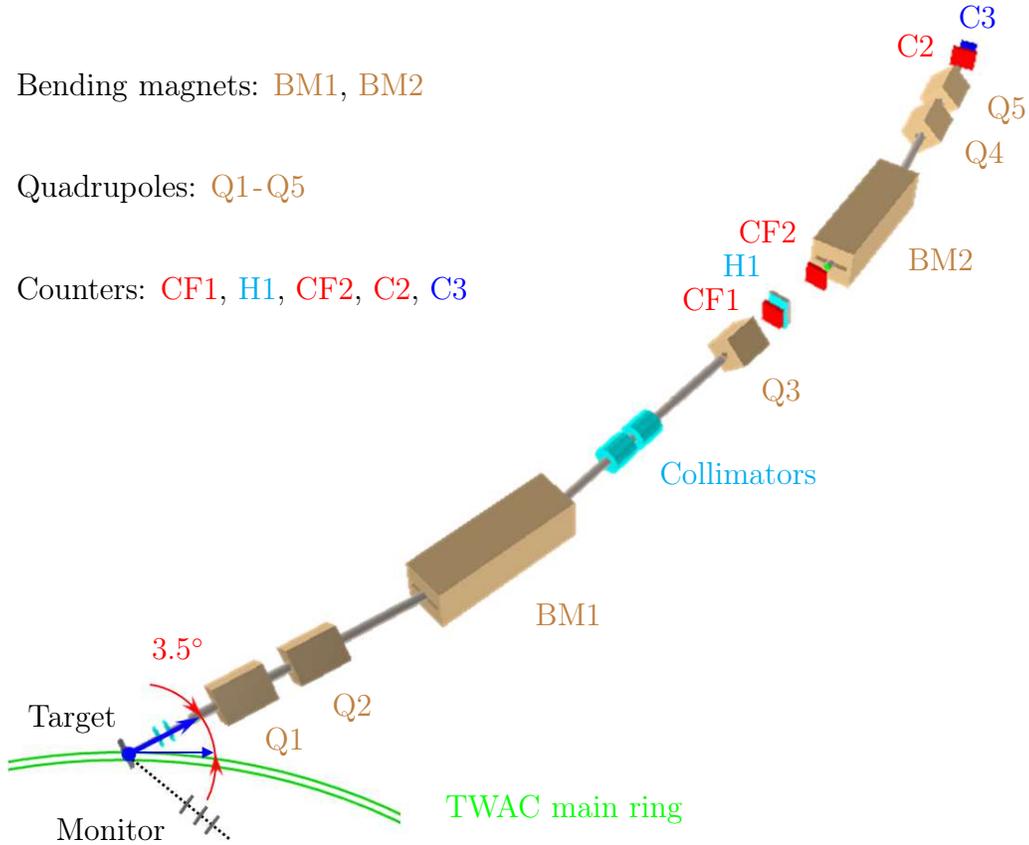
\begin{figure}[!htb]	
\setlength{\unitlength}{1.5mm}
\includegraphics[scale = 0.35]{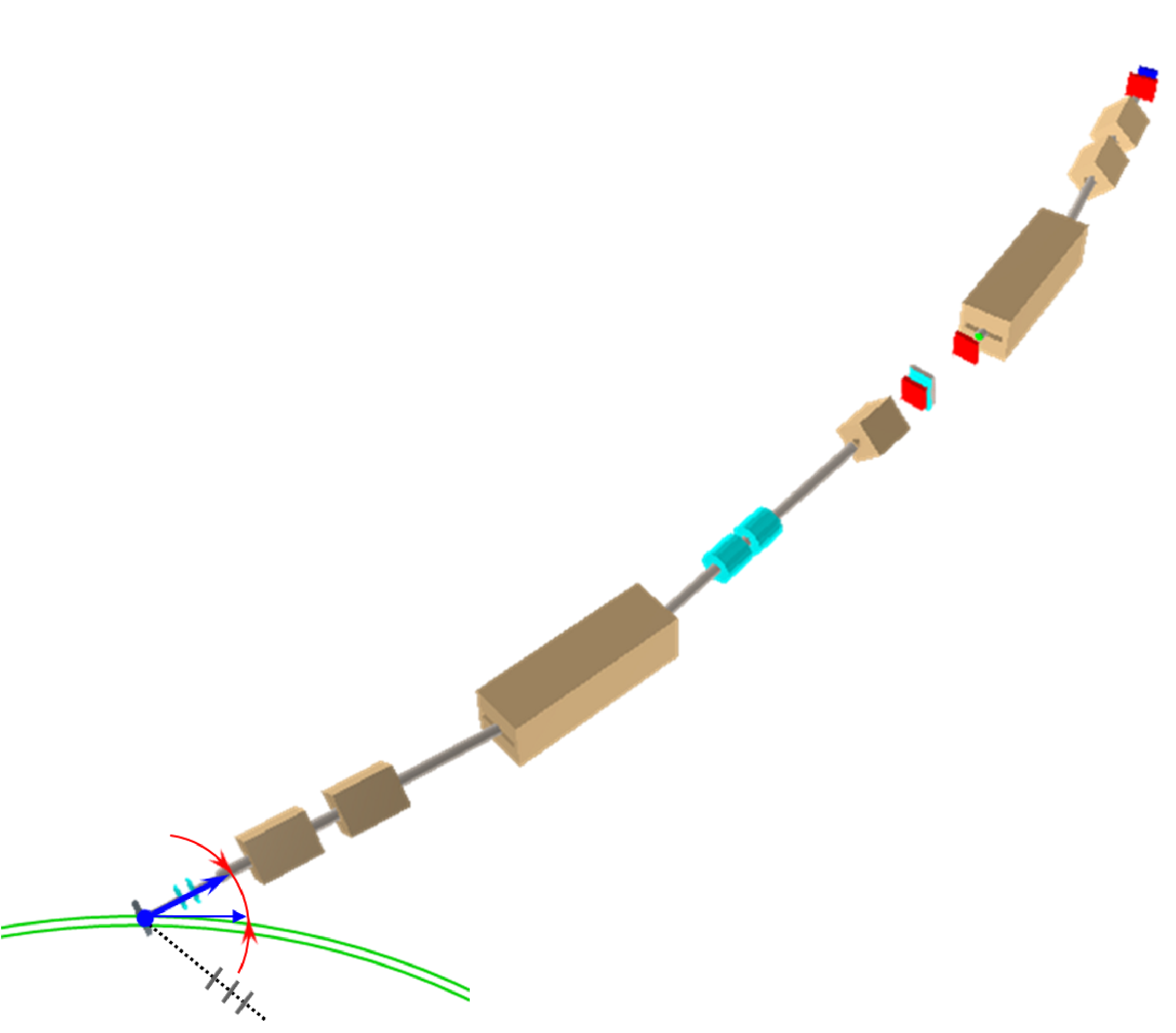}
\begin{picture}(400,0)
\put(13,16){\color{black}\fontsize{12}{12}\selectfont{Target}}
\put(15.5,6){\color{black}\fontsize{12}{12}\selectfont{Monitor}}
\put(24,22){\color{red}\fontsize{12}{12}\selectfont{3.5$^{\circ}$}}
\put(34.,14.0){\color{brown}\fontsize{12}{12}\selectfont{Q1}}
\put(40,17){\color{brown}\fontsize{12}{12}\selectfont{Q2}}
\put(58,25){\color{brown}\fontsize{12}{12}\selectfont{BM1}}
\put(69,37.5){\color{cyan}\fontsize{12}{12}\selectfont{Collimators}}
\put(78,45){\color{brown}\fontsize{12}{12}\selectfont{Q3}}
\put(71,53){\color{red}\fontsize{12}{12}\selectfont{CF1}}
\put(74.6,56){\color{cyan}\fontsize{12}{12}\selectfont{H1}}
\put(76,59){\color{red}\fontsize{12}{12}\selectfont{CF2}}
\put(91,56.5){\color{brown}\fontsize{12}{12}\selectfont{BM2}}
\put(96,66){\color{brown}\fontsize{12}{12}\selectfont{Q4}}
\put(98,70){\color{brown}\fontsize{12}{12}\selectfont{Q5}}
\put(90,75.5){\color{red}\fontsize{12}{12}\selectfont{C2}}
\put(95.5,78){\color{blue}\fontsize{12}{12}\selectfont{C3}}
\put(50,8){\color{green}\fontsize{12}{12}\selectfont{ТWAC main ring}}
\put(12,72){\color{black}\fontsize{12}{12}\selectfont{Bending magnets: 
\color{brown}{BM1}\color{black}{,} \color{brown}{BM2}}}
\put(12,63){\color{black}\fontsize{12}{12}\selectfont{Quadrupoles:
\color{brown}{Q1}\hspace{1pt}\color{brown}-\hspace{1pt}\color{brown}{Q5}}}
\put(12,54){\color{black}\fontsize{12}{12}\selectfont{Counters:
\color{red}{CF1}\color{black}{,} \color{cyan}{H1}\color{black}{,} 
\color{red}{CF2}\color{black}{,} \color{red}{C2}\color{black}{,} \color{blue}{C3}}}
\end{picture}\vspace{-0.8cm}
\caption{Layout of the FRAGM experimental setup~\cite{PaperIntr10}.}\label{Pict02}
\end{figure}

he experimental setup~\cite{PaperSetup5,PaperSetup6} is a two-stage magneto-optical channel of 42~m long oriented at an angle of 3.5$^{\circ}$
relative to the accelerator beam. A narrow vertical beryllium foil with a thickness of 50~$\mu$m was used as a target located in the accelerator vacuum chamber. This made it possible to simultaneously have high luminosity due to multiple passage of ions through the target and small source sizes to fully utilize the high momentum resolution of the channel. Layout of the FRAGM facility is shown in Fig.~\ref{Pict02}.

The first stage of the channel consisted of a doublet of quadrupoles Q1 and Q2, a bending magnet BM1, a system of collimators and a field quadrupole Q3 intended to improve the beam momentum resolution. The second stage consisted of a same doublet of quadrupoles (Q4 and Q5) and a BM2 bending magnet, which directs the beam to area of the second scintillation counters set. The magnetic field in the BM1 and BM2 magnets was controlled through Hall sensors. The vacuum pipeline of the setup has a variable radius from 3.3~cm at the beginning of channel to 10~cm in the rest of it. In the area of first focus, located at a distance of 28~m from the target, vacuum pipeline has a gap in which the first set of scintillation counters CF1 and CF2 were located. The second set of counters C2 and C3 were placed in the area of second focus. The transverse size of CF1, CF2 and C2 is 20$\times$8~cm$^2$, C3 -- 5$\times$5~cm$^{2}$. The scintillation counters were intended for amplitude and time-of-flight measurements.

Each counters were scanned by two photomultipliers from opposite sides in order to compensate their geometric dimensions in time-of-flight measurements~\cite{PaperSetup7}. The coincidence of signals from two counters at different foci was used as a trigger for reading and writing information to the disk. In region of the Q3 lens there was an intermediate focus with a large horizontal dispersion, where the H1 hodoscope was installed. It consisted of twenty vertical elements with size \mbox{20$\times$1$\times$1~cm$^3$} and was used to measure the beam profile and to refine the fragment momentum taking into account the focusing properties of the magneto-optical channel. Initial beam was tested using three scintillation counters directed to the target at an angle of 2$^\circ$.

The data acquisition system of the FRAGM experiment was consisted of three CAMAC crates, where located the modules of time-to-digital and charge-to-digital converters, pulse counters, input registers from hodoscope counters, and output registers. All crate controllers were combined into one CAMAC branch connected via interface boards to a buffer computer with the LINUX operating system. The interface was organized according to the built-in memory scheme, which allowed the entire address space of the CAMAC branch to be transferred directly into the address space of the computer memory. In addition to this convenience of  data access and programming, this scheme ensured the maximum readout rate for the CAMAC system of 1~MHz with 16-bit words. The readout time was less than 50~$\mu$s, which allowed up to 10$^4$ events to be collected per accelerator reset with duration about one second. In the intervals between cycles, data was send to the main computer via a fast communication line. The control of data collection, the monitoring of the experiment progress and the recording to disk were performed using the main computer. During the data collection operator has the possibility to control such parameters as the monitor indications, the amplitude-time information received from the counters and to analyze the ions most typical for the specific magneto-optical channel setting. The software for both the experiment control and the preliminary data analysis was created on the basis of the ROOT~\cite{PaperSetup8} software package. At the end of the data acquisition all the information was copied to the main experiment data storage server via the local institute network.

The data acquisition to obtain the experimental material for studying charge-exchange reactions was carried out on a beam of carbon ions at energy of 300~MeV/nucleon. In this case, the magneto-optical channel was adjusted to a certain rigidity in the range from 0.9 to 2.8~GeV/{\it c}. At 
At special adjustments, especially in the region from 1.7 to 2.4~GeV/{\it c}, which are characteristic of reactions occurring with nucleon charge-exchange, the data acquisition were carried out several times in order to obtain a larger volume of experimental data. The integral value of the monitor over entire period of the data collection is \mbox{$\sim$~5.4$\times$10$^8$} counts, which corresponds to \mbox{$\sim$~1.5$\times$10$^7$} events of nucleus-nucleus interactions detected by the FRAGM detector.

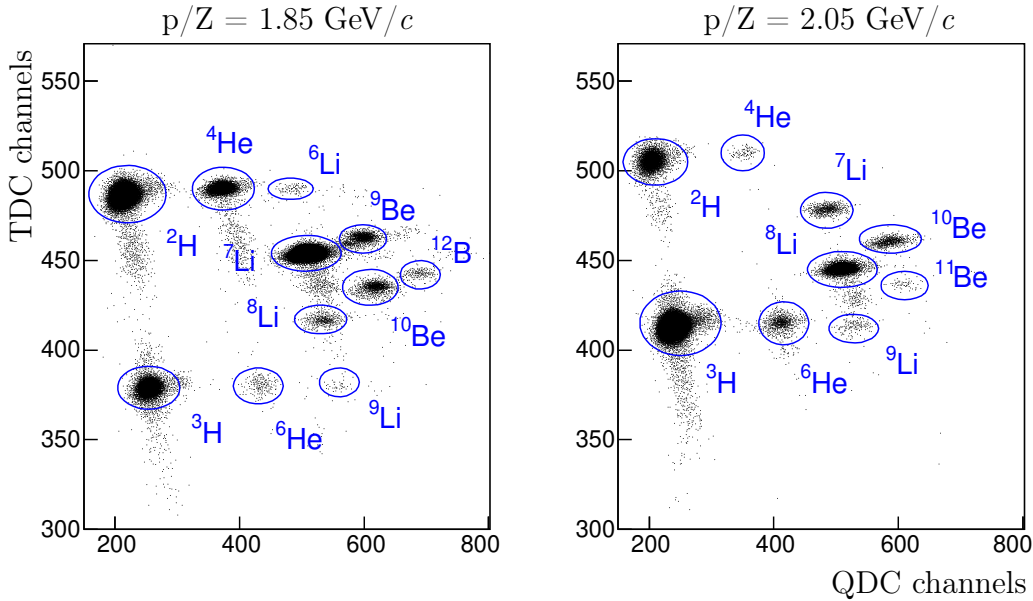
\begin{figure}[htp]\vspace{0.5cm}
\includegraphics[scale=0.74]{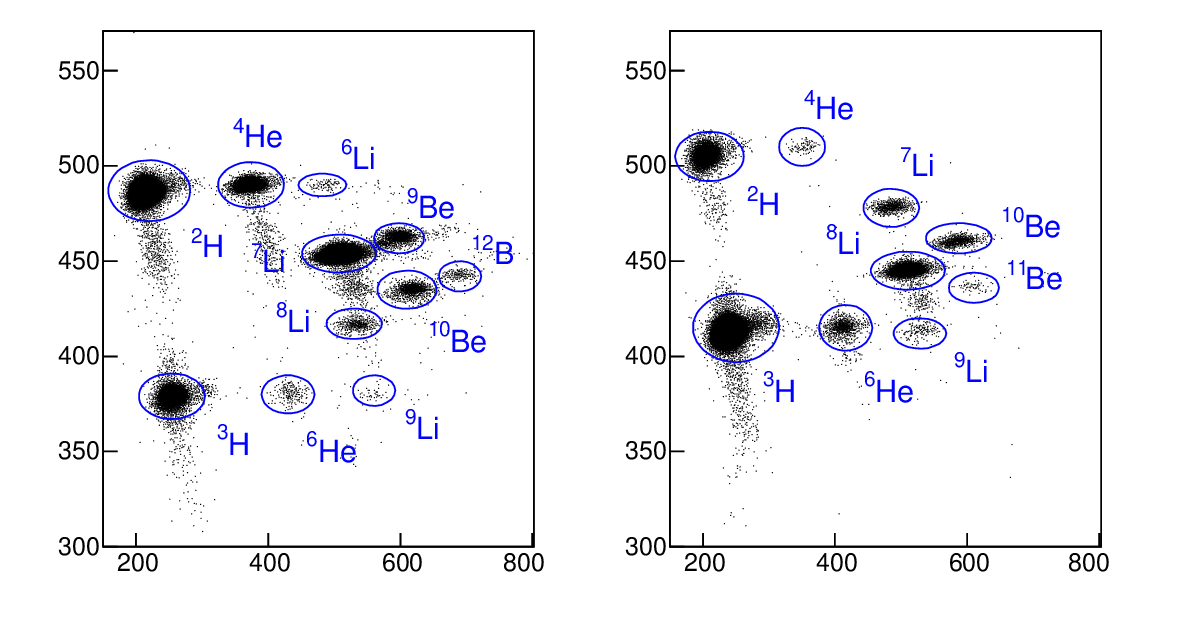}
\setlength{\unitlength}{1.5mm} 
\begin{picture}(400,0)
\put(79,5.5){\fontsize{12}{12}\selectfont\hspace{0.2cm}{QDC channels}}
\put(7.5,36.5){\rotatebox{90}{\fontsize{12}{12}\selectfont{TDC channels}}}
\put(21,55.5){\fontsize{12}{12}\selectfont{p/Z = 1.85 GeV/\it{с}}}
\put(69,55.5){\fontsize{12}{12}\selectfont{p/Z = 2.05 GeV/\it{с}}}
\end{picture}\vspace{-0.8cm}
\caption{Correlation distribution of the time-of-flight (TDC channels) and signal amplitude (QDC channels) for two 
rigidities of the magneto-optical channel of the FRAGM experiment.}\label{Pict03}
\end{figure}

The process of selection the desired fragment consisted of analyzing the correlation distributions of the time-of-flight and amplitude in the scintillation counter CF1. The time-of-flight, which is measured between the counters CF1 and C2, is proportional to the fragment mass number. The amplitude is proportional to the energy losses in the scintillator and is a function of the registered fragment charge. A typical picture of such distributions is shown in Fig.~\ref{Pict03} for two rigidities of the magneto-optical channel. 

As the rigidity decreases, the ions of same charge are shifting to the region of higher amplitudes, which is associated with a decrease in their velocity. In addition, the correlation distributions contain background events are caused by both the simultaneous entry of two particles into the scintillator and interaction of fragments with the elements of magneto-optical channel and scintillation counters. It is seen, that the signals from different fragments are well separated. In particular, at the rigidity of 1.85~GeV/{\it c}, 11 fragments can be identified, including the $^{12}$B fragment obtained as a result of the nucleon charge-exchange reaction. At the rigidity of 2.05~GeV/{\it c}, 9 fragments can be identified. Measurements of the ratio of lithium, beryllium, and boron isotopes yields to the yield of stable $^{4}$He ion are a stability markers when collecting several runs with the same rigidity.

When analyzing the correlation distributions, it is evident that the signals from fragments with the same charge are shifted in amplitude (QDC channels) relative to each other. This is due to that this amplitude is proportional to the ionization losses in the scintillator, which are increased with decreasing fragment velocity and also with an increase of the fragment mass at a fixed rigidity. The transition from amplitude to charge will remove that shift and it simplifies the analysis of correlation distributions. An important parameter of any scintillator is a light yield determining the signal amplitude. Light yield of a scintillator depends on its production material, on the ratio of emission and absorption spectra, on the production technology, and on the type of particle that causes scintillations. Birks' empirical law describing the light yield per path length ($dL/dx$) as a function on the particle energy losses $dE/dx$ when it passes through the scintillator, is determined by the formula: $dL/dx~=~SdE/dx/(1+k_{B}dE/dx)$, where $S$ is the scintillation efficiency, $k_B$ is the Birks coefficient, which depends only on the properties of scintillator~\cite{PaperSetup9,PaperSetup10}. Based on experimental data for a wide range of fragments from protons to carbon, ionization losses $dE/dx$ were calculated using the Bethe-Bloch formula. Dependence of the light yield $dL/dx$ on the amplitude of measured signal is nonlinear and can be well described by the 3rd degree polynomial. This dependence was used to determine the charge at arbitrary values of the amplitude and time-of-flight. Thus, the procedure for such recalculating into a charge allows aligning the signals for isotopes of one element, which simplifies the procedure for isolating the desired fragment. In addition, this recalculation allows to improve the resolution when amplitude is converted to charge.

\subsection{Simulation of ion beam passage in a magneto-optical channel. }

The magneto-optical channel of the FRAGM setup has special construction features. Charged particles moving along the channel, pass through collimators, scintillation counters, vacuum line breaks in the counter area, which affects the beam intensity and energy. Obviously, when transporting the beam, it is necessary to take into account such processes as multiple scattering, ionization losses, and inelastic interaction of particles in the matter. Moreover, it is important to determine the momentum and angular capture of the setup using in the physical analysis of experimental data. To solve such problems, a simulation program was created using the Geant4~\cite{PaperSetup11} software package. The program code included a precise description of the geometric parameters of the magneto-optical channel elements and scintillation counters, measured maps of the magnetic fields of the bending magnets and quadrupoles. Magnetic fields and quadrupole gradients were set in accordance with the channel adjustment to a certain rigidity value ($p$/$Z$). The main parameters of the channel were determined for protons with a momentum of 1~GeV/{\it c} during the particle transportation from the target region to second focus. The angular capture of the experimental setup was \mbox{$\Delta\theta$~$\approx$~$\pm$~0.5$^{\circ}$} at an angle of 3.5$^{\circ}$, momentum capture -- \mbox{$\Delta p/p$~$\approx$~3.5\%}.

\begin{figure}[!htb]\vspace{0.2cm}
\includegraphics[scale=0.56]{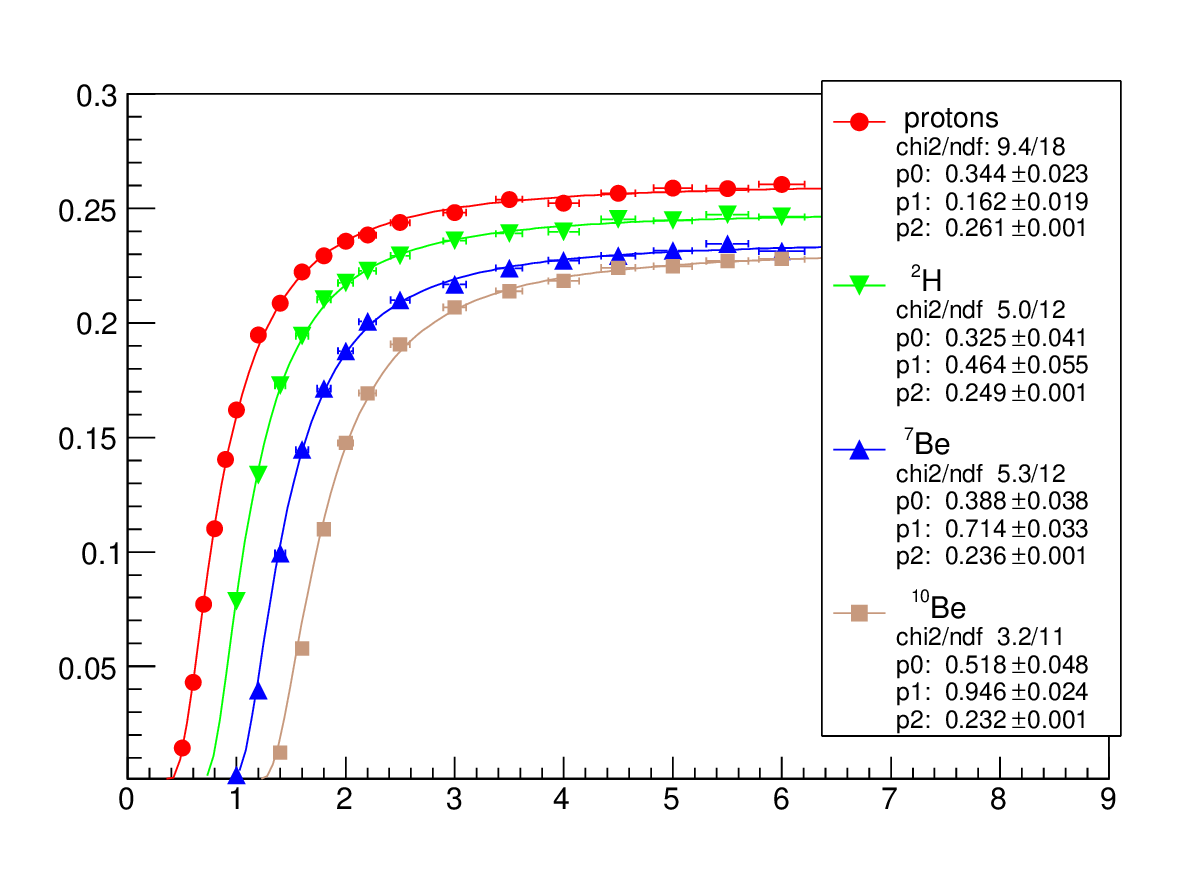}
\setlength{\unitlength}{1.5mm} 
\begin{picture}(400,0)
\put(72.5,6.2){\fontsize{12}{12}\selectfont{$p$/$Z$, GeV/{\it c}}}
\put(18.,43.){\rotatebox{90}{\fontsize{12}{12}\selectfont{Efficiency}}}
\end{picture}\vspace{-1.2cm}
\caption{Proton and ion detection efficiency by the magneto-optical channel of the FRAGM setup depending on the channel rigidity~\cite{PaperIntr10}.}\label{Pict04}
\end{figure}
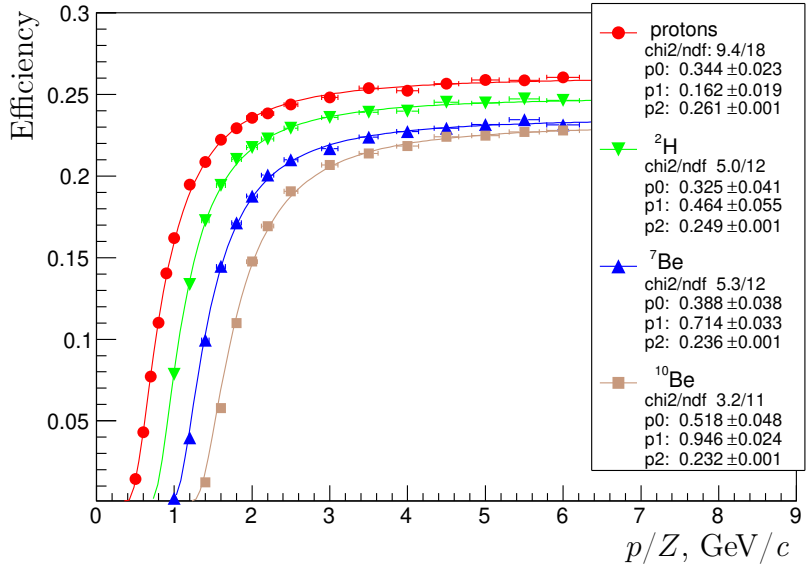
The detection efficiency of various ions by the magneto-optical channel depending on it's momentum was determined in a wide range of rigidity 
from 0.6 to 6~GeV/{\it c}. The charged particle is detected if it passes through the counters in both the first and second focus. This detection efficiencies for protons, deuterons and beryllium isotopes ($^7$Be and $^{10}$Be) are shown in Fig.~\ref{Pict04}. The model data can be well described by the function 
\mbox{$f(x)~=~p_{2}\times~exp[-p_{0}/(x~-~p_{1})^{2}]$}, where $x$~=~$p$/$Z$, $p$ and $Z$ are the fragment momentum and charge, $p_{0}$, $p_{1}$, $p_{2}$ are free parameters. The efficiency correction plays an important role at rigidity up to 3~GeV/{\it c} and can significantly 
affect the fragment momentum spectrum. In the region of large rigidity values, the proton and ion efficiencies reach a plateau, and for protons it is approximately 20\% greater than for beryllium.

\section{DIFFERENTIAL CROSS SECTIONS OF REACTIONS WITH NUCLEON CHARGE-EXCHANGE}

\subsection{Model description of nucleus-nucleus interactions. }

As noted earlier, computer tools based on both the most modern theoretical approaches and experimental data are used as calculation tools for modeling nucleus-nucleus interaction reactions. The applicability limits of a particular model are determined by both the energy of projectile nucleus and the masses of colliding nuclei. Statistical models are usually used to describe interactions occurring at intermediate energies; the intranuclear cascade model is widely used. The corresponding theoretical predictions give reasonable results in the energy range up to several GeV and for relatively light nuclei, for which multiparticle interaction processes can be neglected. The most important task in testing models of nucleus-nucleus collisions is their correspondence to experimental data, both in the widest possible range of energies and in the angular dependence of the fragment yields. Undoubtedly, an increase of the volume of experimental data should improve the models of nucleus-nucleus interactions.

In this study three models of nucleus-nucleus interactions were investigated in terms of their agreement with our experimental data obtained for charge-exchange reactions during fragmentation of carbon nuclei. The following models were studied: Binary Cascade (BC)~\cite{PaperBC}, Liege Intranuclear Cascade (INCL)~\cite{PaperINCL}, Quantum Molecular Synamics (QMD)~\cite{PaperQMD}. Some features of these models should be noted. The BC model, based on the principles of classical intranuclear cascade, has a significant limitation on the mass of interacting nuclei and is in good agreement
with experimental data only for light nuclei. QMD is a quantum continuation of the classical molecular dynamics model and usually used to describe nuclear interactions with heavy nuclei. The INCL model was created to describe the interaction of a carbon nuclei with nuclear matter at intermediate energies, which is important, in particular, for the purposes of medical physics. As in the case of BC, this model has restrictions on the mass of nuclei and gives a more accurate description for light nuclei.

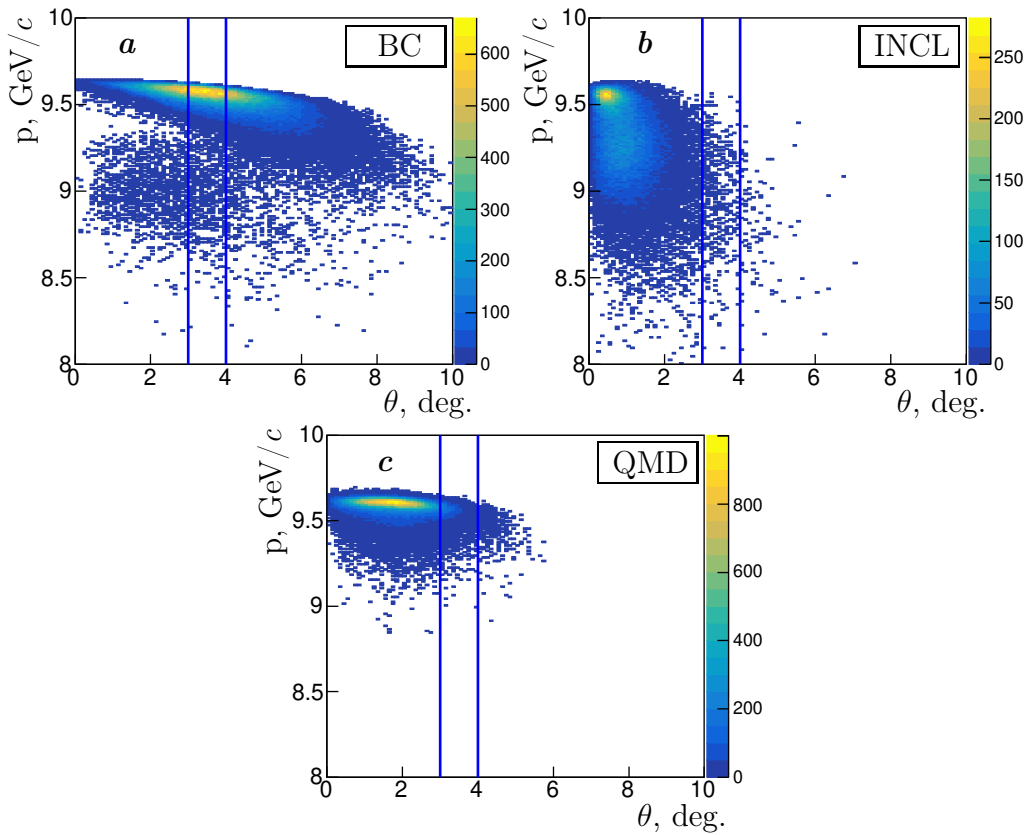
\begin{figure}[!htb]
\includegraphics[scale=0.67]{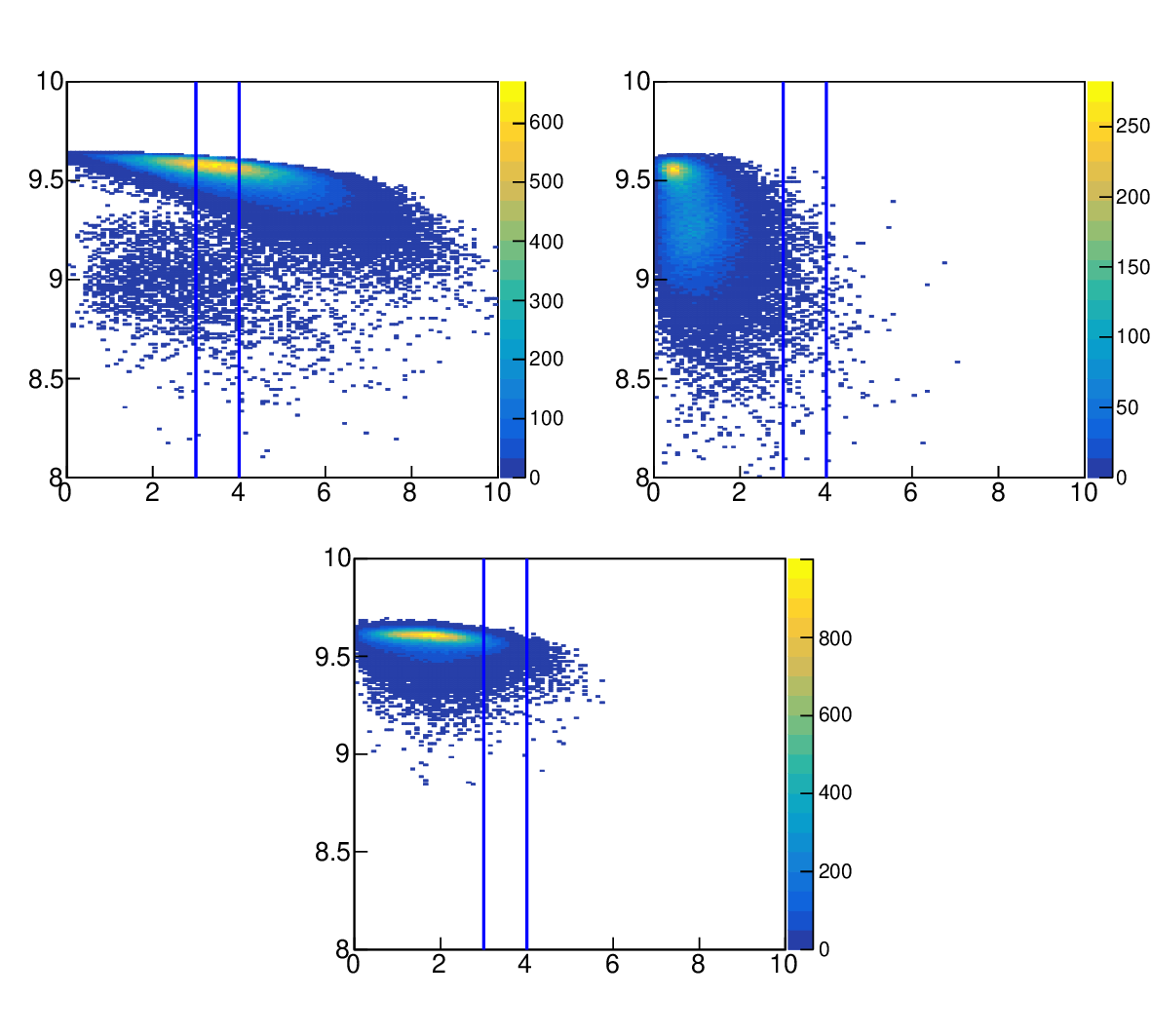}
\setlength{\unitlength}{1.5mm} 
\begin{picture}(400,0)
\put(9.4,68){\rotatebox{90}{\fontsize{12}{12}\selectfont{p, GeV/{\it c}}}}
\put(55,68){\rotatebox{90}{\fontsize{12}{12}\selectfont{p, GeV/{\it c}}}}
\put(62.5,7.5){\fontsize{12}{12}\selectfont\hspace{0.4cm}{$\theta$, deg.}}
\put(85.5,44.2){\fontsize{12}{12}\selectfont\hspace{0.4cm}{$\theta$, deg.}}
\put(40,44.2){\fontsize{12}{12}\selectfont\hspace{0.4cm}{$\theta$, deg.}}
\put(31.5,31.5){\rotatebox{90}{\fontsize{12}{12}\selectfont{p, GeV/{\it c}}}}
\put(19,76){\fontsize{12}{12}\selectfont{{\it \textbf{a}}}}
\put(65,76){\fontsize{12}{12}\selectfont{{\it \textbf{b}}}}
\put(42,39){\fontsize{12}{12}\selectfont{{\it \textbf{c}}}}
\put(39.3,76.0){\fbox{\parbox[][9pt][t]{1.1cm}{\fontsize{13}{13}\selectfont{\hspace{0.1cm}BC}}}}
\put(84.8,76.0){\fbox{\parbox[][9pt][t]{1.1cm}{\fontsize{13}{13}\selectfont{\hspace{0.085cm}INCL}}}}
\put(61.6,39.05){\fbox{\parbox[][9pt][t]{1.1cm}{\fontsize{13}{13}\selectfont{\hspace{0.085cm}QMD}}}}
\end{picture}\vspace{-0.8cm}
\caption{	
Momentum as a function of angle for $^{12}$B produced in a charge-exchange reaction for three models of nucleus-nucleus interactions: BC ({\it a}), INCL ({\it b}) and QMD ({\it c}).}\label{Pict05}
\end{figure}

Description of charge-exchange processes within the framework of nucleus-nucleus interaction models is illustrated by Fig.~\ref{Pict05} presenting the dependencies of momentum ($p$) from emission angle ($\theta$) of the fragment $^{12}$B~\cite{PaperFRAGM1}. It is evident that the correlation distributions in different models differ significantly
from each other. The vertical lines in the plots show an angular capture region of the FRAGM experimental setup. The BC
model shows the highest yield of $^{12}$B in the region of experimental setup, whereas the yields of this isotope in the INCL and QMD models are significantly suppressed in this region. Comparison of model predictions and experimental data demonstrated that the BC model gives the best description of the momentum shape of $^{12}$B. It should also be noted, that the BC and QMD models give similar shapes of the momentum spectra for $^{12}$B and $^{12}$N. Comparison of the momentum distributions for light ions produced in the typical process of $^{12}$C fragmentation showed that the BC model gives the best agreement with our experimental data. Later, as a normalization factor, we will use the $^4$He yield calculated in the BC model.

\subsection{Identification of fragments on the FRAGM magnetic spectrometer. }

To identify beryllium and boron isotopes, the process of ion selection is based on the analysis of correlation distributions between the time-of-flight and amplitude in the rigidity range from 1 to 2.8~GeV/{\it c} by step of 
50~MeV/{\it c}. The ion separation is carried out using two lines with the same slope, then the selected data are projected onto the time axis (TDC). Fig.~\ref{Pict06} shows the correlation distributions of the time-of-flight by  amplitude and the its projection onto the time axis in the registration region of the $^{10}$Be isotope. It's 
clear seen, that the signals from different ions are well separated, and the number of registered events is determined by the sum of events in the distribution. Background from neighboring isotopes is negligible and not visible on the 
plot. To calculate a differential cross-sections, such factors as monitor readings are taken into account too,
\begin{figure}[!htb]
\includegraphics[scale=0.70]{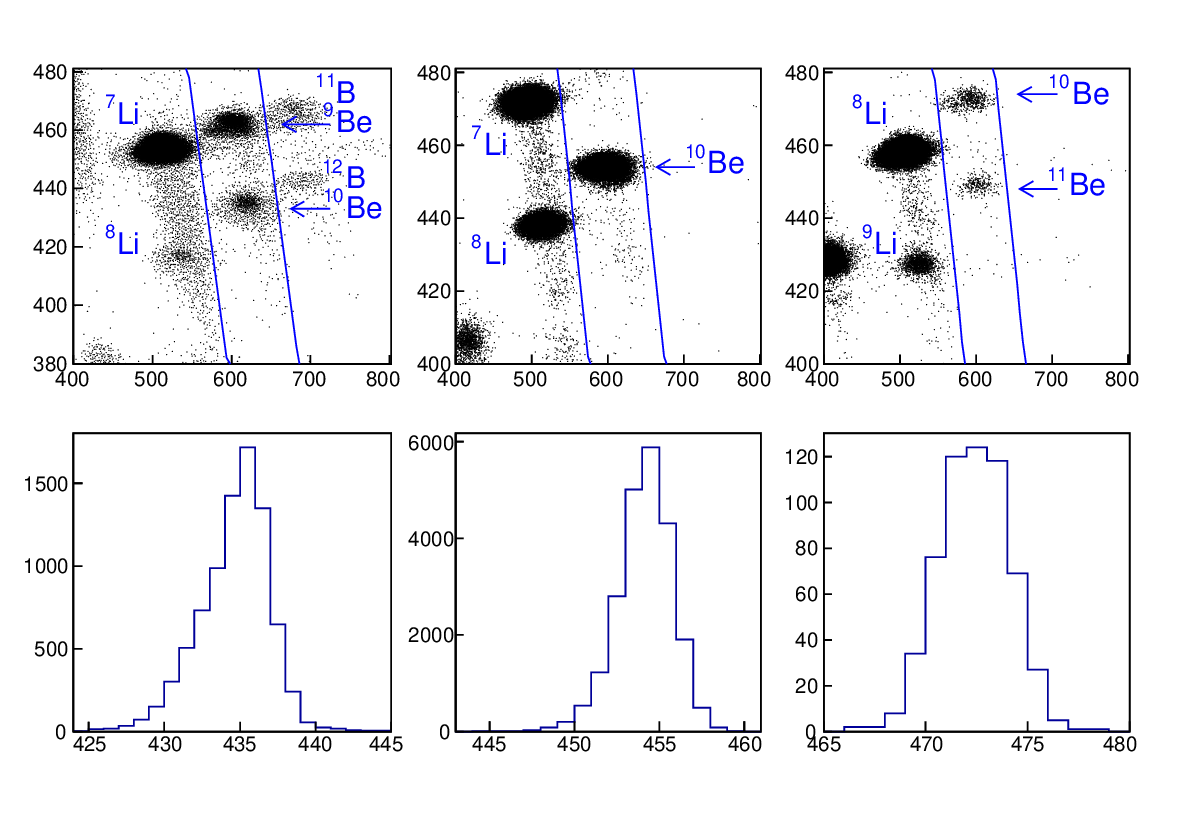}
\setlength{\unitlength}{1.5mm} 
\begin{picture}(400,0)
\put(79.0,8.5){\fontsize{12}{12}\selectfont\hspace{0.2cm}{QDC channels}}
\put(8.0,48){\rotatebox{90}{\fontsize{12}{12}\selectfont{TDC channels}}}
\put(14.8,66.0){\fontsize{11}{11}\selectfont\hspace{0.2cm}{p/Z = 1.85 GeV/{\it c}}}
\put(45.5,66.0){\fontsize{11}{11}\selectfont\hspace{0.2cm}{p/Z = 2.0 GeV/{\it c}}}
\put(74.0,66.0){\fontsize{11}{11}\selectfont\hspace{0.2cm}{p/Z = 2.15 GeV/{\it c}}}
\end{picture}\vspace{-1.2cm}
\caption{Upper row: correlation distributions of the time-of-flight (TDC channels) and signal amplitude (QDC channels) for different settings of the magneto-optical channel for different rigidities. Lower row: projection onto the time axis the distribution of the isotope in the detection region of $^{10}$Be~\cite{PaperIntr10}.}\label{Pict06}
\end{figure}
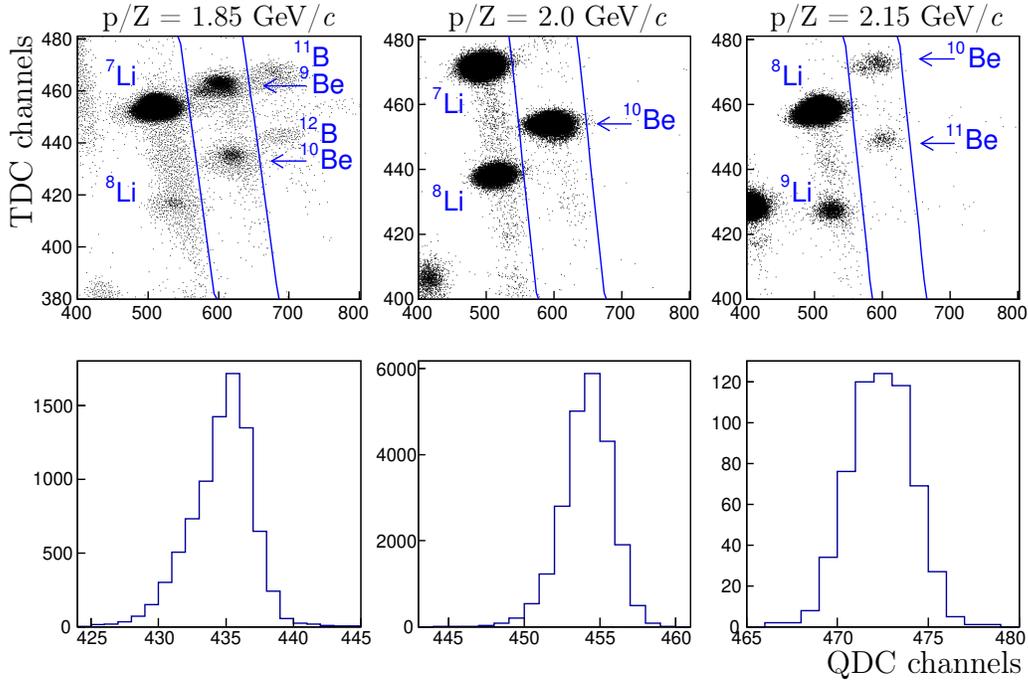
\begin{figure}[!htb]\vspace{0.2cm}
\includegraphics[scale=0.80]{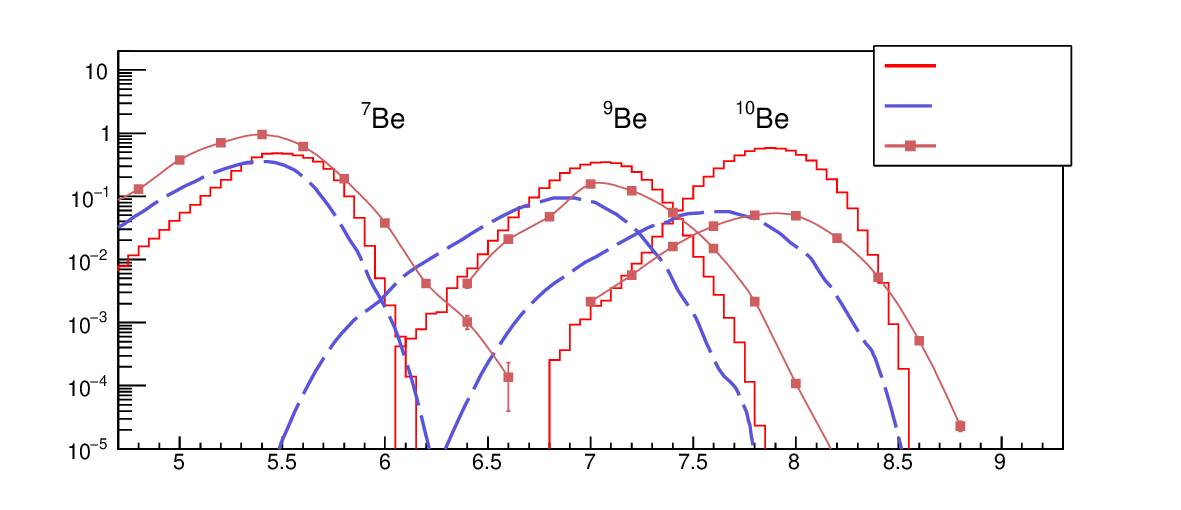}	
\setlength{\unitlength}{1.5mm} 
\begin{picture}(400,0)
\put(84,6.0){\fontsize{12}{12}\selectfont\hspace{0.2cm}{{\it p}, GeV/{\it c}}}
\put(24,42){\fontsize{14}{14}\selectfont{{\it \textbf{a}}}}
\put(13,48.0){\fontsize{12}{12}\selectfont{{\it d}$^{\hspace{0.1em}2}\sigma/$({\it dpd}$\Omega$), 
mb$/$(MeV$/{\it с}\cdot$\hspace{-0.1cm} sr)}}
\put(88,44.6){\fontsize{9}{9}\selectfont{BC}}
\put(88,41.1){\fontsize{9}{9}\selectfont{INCL}}
\put(88,37.3){\fontsize{9}{9}\selectfont{FRAGM}}	
\end{picture}
\includegraphics[scale=0.80]{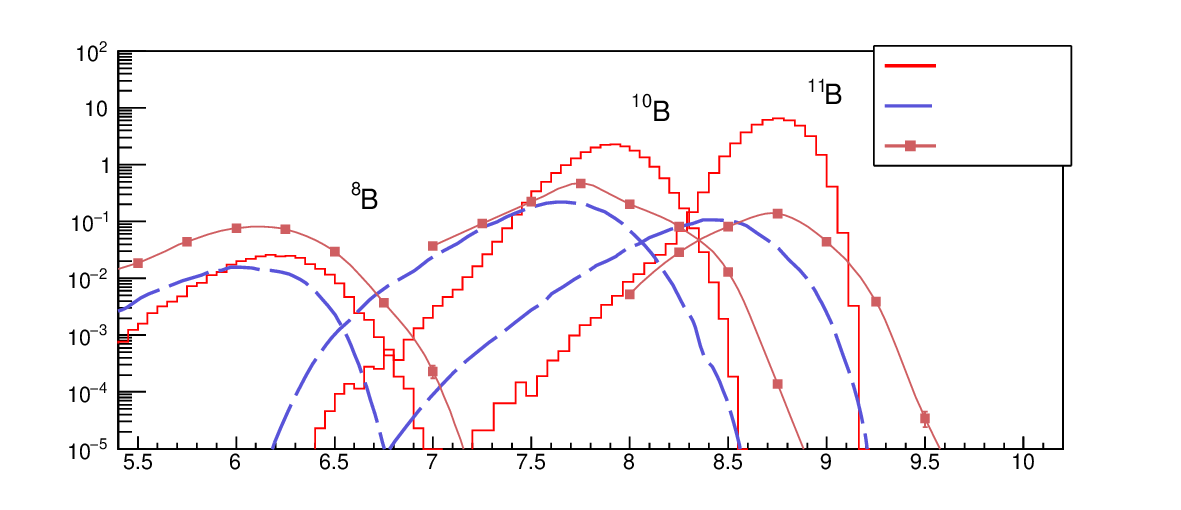}	
\begin{picture}(400,0)	
\put(84,6.0){\fontsize{12}{12}\selectfont\hspace{0.2cm}{{\it p}, GeV/{\it c}}}
\put(24,40){\fontsize{14}{14}\selectfont{{\it \textbf{b}}}}
\put(13,48.0){\fontsize{12}{12}\selectfont{{\it d}$^{\hspace{0.1em}2}\sigma/$({\it dpd}$\Omega$), 
mb$/$(MeV$/{\it с}\cdot$\hspace{-0.1cm} sr)}}
\put(88,44.6){\fontsize{9}{9}\selectfont{BC}}
\put(88,41.1){\fontsize{9}{9}\selectfont{INCL}}
\put(88,37.3){\fontsize{9}{9}\selectfont{FRAGM}}	
\end{picture}\vspace{-0.8cm}
\caption{Differential cross sections of beryllium ({\it a}) and boron ({\it b}) ion yields as a function of
momentum~\cite{PaperFRAGM1}.}\label{Pict07}
\end{figure}
total interaction cross-section and ion detection efficiency of the magneto-optical channel of the FRAGM experiment.

The total cross section of carbon and beryllium ions interaction via the inelastic channel $\sigma_{tot}$ can be determined using the empirical parameterization~\cite{PaperSihver}. This approximation is energy-independent and the calculated cross section gives \mbox{$\sigma_{tot}$~=~776.8~mb}. A more detailed calculation of the total cross section depending on the projectile ion energy was calculated using the LAQGSM model~\cite{PaperLAQGSM}. The cross section value at an energy of 300~MeV/nucleon obtained by the model is 772.8~mb, which coincides by 0.5\% with the cross section value calculated by formula. On the other hand, the difference between these predictions reaches a level of 10\% at an energy of 1~GeV/nucleon. Later the value of total cross section calculated by the LAQGSM model will be used to normalize the experimental data.

Fig.~\ref{Pict07} shows the measured and model data for differential cross sections as a function of the laboratory momentum for the boron and beryllium isotopes. The distribution shapes of each fragment are looking like a Gaussian distribution, where the position of momentum maximum is close to the corresponding momentum per nucleon of the projectile carbon nucleus. The simulation of experiment was performed using the BC and INCL nucleus-nucleus interaction models. The difference in absolute cross section is due to significant differences between the predictions of ion-ion interaction models. The models have various approaches to describing these interactions, which leads to differences in the cross section values. However, there is good agreement between the model and measured data in terms of the mean value and the shape of distribution, which indicates the similarity of models in describing physical phenomena, despite the differences in absolute cross section values.

\subsection{Differential cross sections of $^{11}$B and $^{12}$B produced in a single nucleon charge-exchange. }
	
\begin{figure}[!htb]\vspace{0.2cm}	
\includegraphics[scale=0.69]{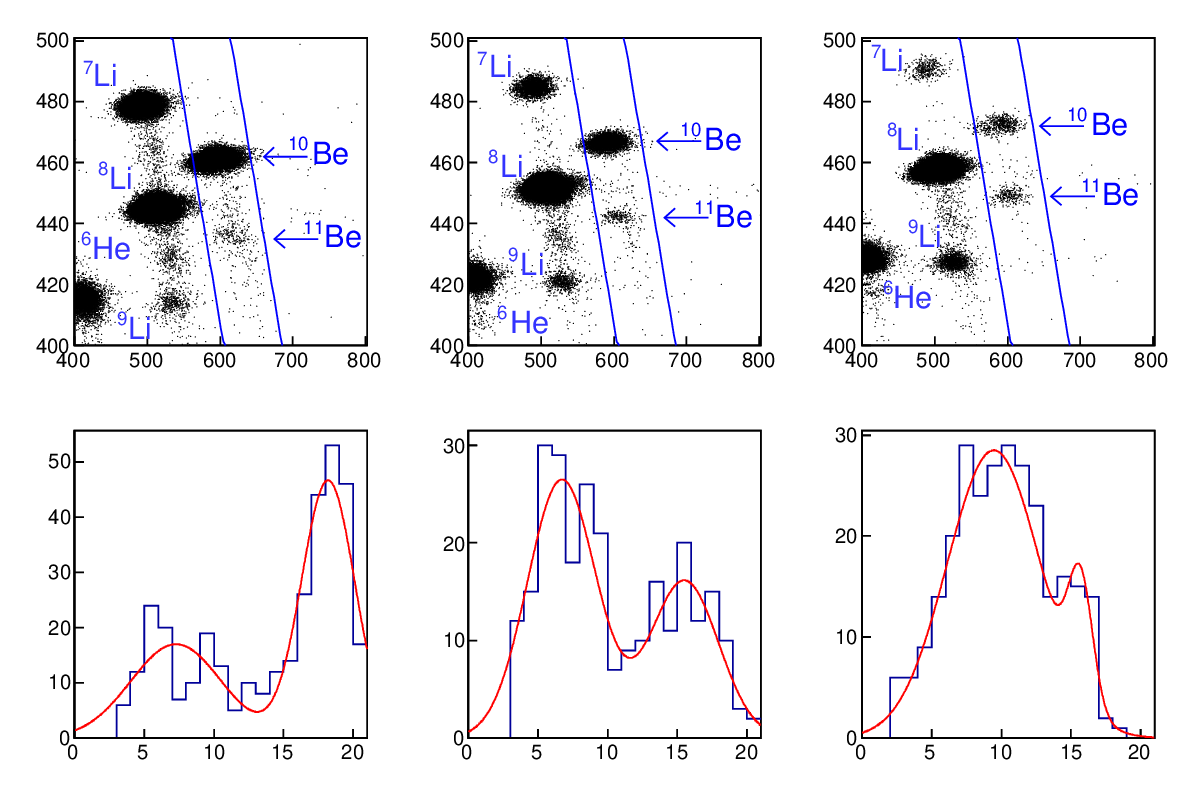}
\setlength{\unitlength}{1.5mm} 
\begin{picture}(400,0)
\put(80.0,35.8){\fontsize{12}{12}\selectfont\hspace{0.2cm}{QDC channels}}
\put(81.0,5.5){\fontsize{12}{12}\selectfont{{\it N} (hodoscope)}}
\put(9.0,47.5){\rotatebox{90}{\fontsize{12}{12}\selectfont{TDC channels}}}
\put(14.8,66.0){\fontsize{11}{11}\selectfont\hspace{0.2cm}{p/Z = 2.05 GeV/{\it c}}}
\put(45.5,66.0){\fontsize{11}{11}\selectfont\hspace{0.2cm}{p/Z = 2.10 GeV/{\it c}}}
\put(76.0,66.0){\fontsize{11}{11}\selectfont\hspace{0.2cm}{p/Z = 2.15 GeV/{\it c}}}
\end{picture}\vspace{-0.8cm}
\caption{Upper row: correlation distributions of time-of-flight (TDC channels) and signal amplitude (каналы QDC) 
for different settings of the magneto-optical channel in terms of rigidity (p/Z). Lower row: distribution of events vs hodoscope cells selected according to the correlation distribution in the area of $^{11}$Be isotope registration.}\label{Pict08}
\end{figure}
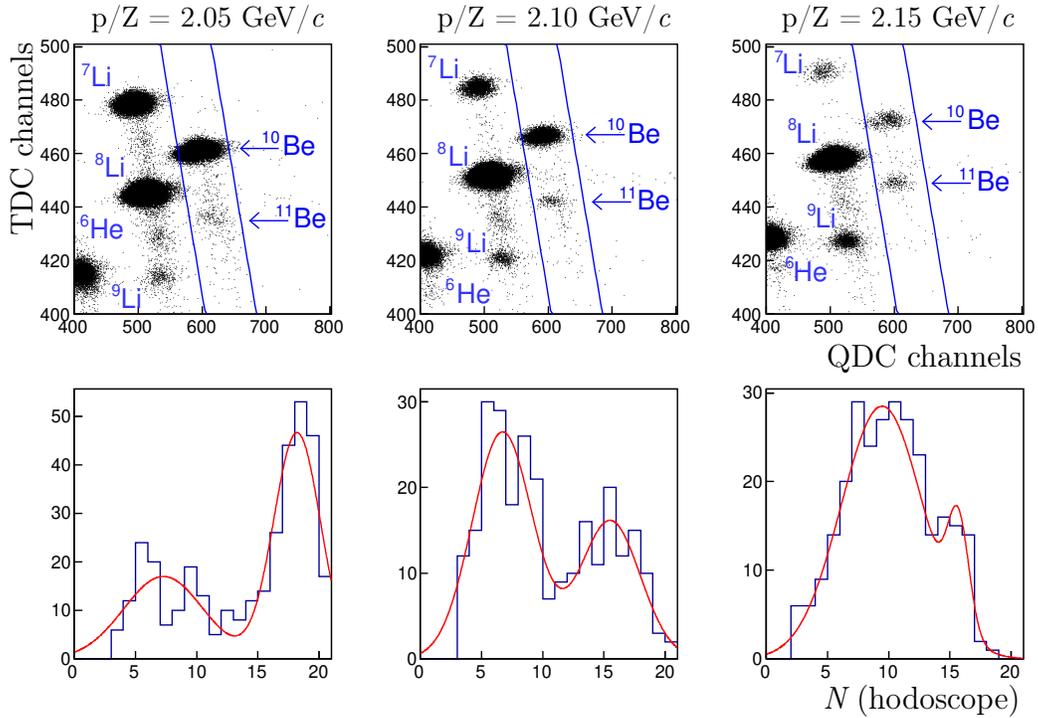

As a result of $^{12}$C ion fragmentation, three isotopes can be produced, due to nucleon charge-exchange:
$^{11}$Bе (7 neutrons), $^{12}$B (7 neutrons) and $^{12}$N (7 protons). The rigidity range from 2.0 to 2.25~GeV/{\it c} was chosen to search  $^{11}$Be isotope. The corresponding correlation distributions of the time-of-flight and charge for different adjustments of the magneto-optical channel are shown in Fig.~\ref{Pict08} in the upper part. Selected candidates for the correlation distributions are analyzed by the hodoscope cells (Fig.~\ref{Pict08}, lower plots). The distribution of such events displays two maxima: the left one corresponds to $^{11}$Be, the right one to the background events from $^{10}$Be. At low rigidities there is a significant contribution from the background. After subtraction of the background part from the histogram, then it corresponds to the signal from $^{11}$Be.
	
Reactions with single nucleon charge-exchange have significantly smaller cross sections and are distinguished by small peak widths in their momentum distributions compared to simple fragmentation processes. Since the measurements were carried out by step of 50~MeV/{\it с}, then this does not give us the opportunity to measure the momentum distribution. Therefore, for a given rigidity, it is necessary to analyze the fragment beam profile in the first focal plane, which can be obtained by using a hodoscope counter. For fixed rigidity, the hodoscope allows to obtain at least 10 additional measurements and refine the fragment momentum to 0.4\%. 
	
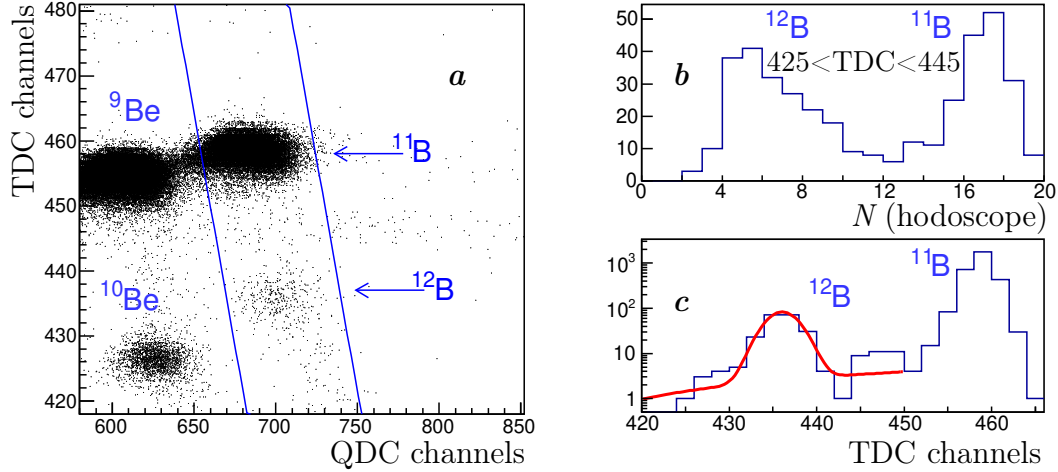
\begin{figure}[!htb]
\includegraphics[scale=0.76]{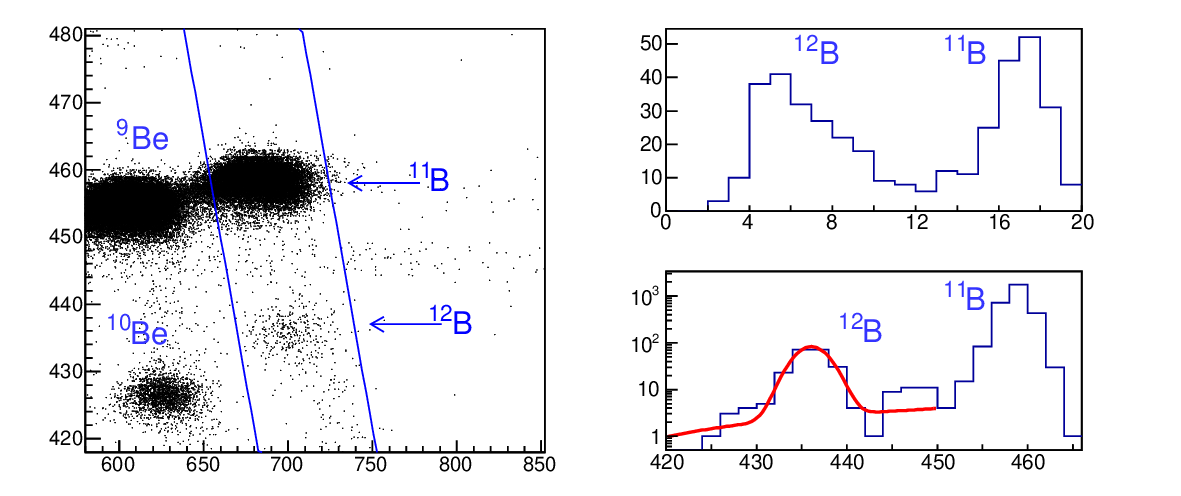}
\setlength{\unitlength}{1.5mm} 
\begin{picture}(400,0)	
\put(35,4.5){\fontsize{12}{12}\selectfont\hspace{-0.2cm}{QDC channels}}
\put(80.5,25.5){\fontsize{12}{12}\selectfont\hspace{-0.1cm}{{\it N} (hodoscope)}}
\put(5.5,28){\rotatebox{90}{\fontsize{12}{12}\selectfont\hspace{0.0cm}{TDC channels}}}
\put(79.6,4.5){\fontsize{12}{12}\selectfont{TDC channels}}
\put(44,38){\fontsize{13}{13}\selectfont{{\it \textbf{a}}}}
\put(64,38){\fontsize{13}{13}\selectfont{{\it \textbf{b}}}}
\put(72.4,39.3){\fontsize{11}{11}\selectfont{425<TDC<445}}
\put(64,18){\fontsize{13}{13}\selectfont{{\it \textbf{c}}}}	
\end{picture}\vspace{-0.5cm}
\caption{
Selection algorithm for $^{12}$B: time-of-flight (TDC) and signal amplitude (QDC) correlation distributions ({\it a}); hodoscope ({\it b}) and TDC channels ({\it c}) distributions obtained by events selected from the corresponding region of correlation distribution~\cite{PaperFRAGM1}.}\label{Pict09}
\end{figure}

The search for fragments corresponding to the $^{12}$B isotope was carried out within the rigidity range from 1.75 to 1.95~GeV/{\it c}. The $^{12}$B isotope selection algorithm is similar to the $^{11}$Be isotope search. For a channel rigidity of 1.8~GeV/{\it c}, the correlation distribution is shown in Fig.~\ref{Pict09} ({\it a}). The distribution over hodoscope cells is shown in Fig.~\ref{Pict09}~({\it b}). The background events produce $^{11}B$ ions, the contribution from which was determined using the fitting procedure. Plot~\ref{Pict09}~({\it c}) presents the distribution of all selected $^{12}$B candidates, summed over all rigidity values. The difference in maxima of the signal peak and background one is of the same order. 
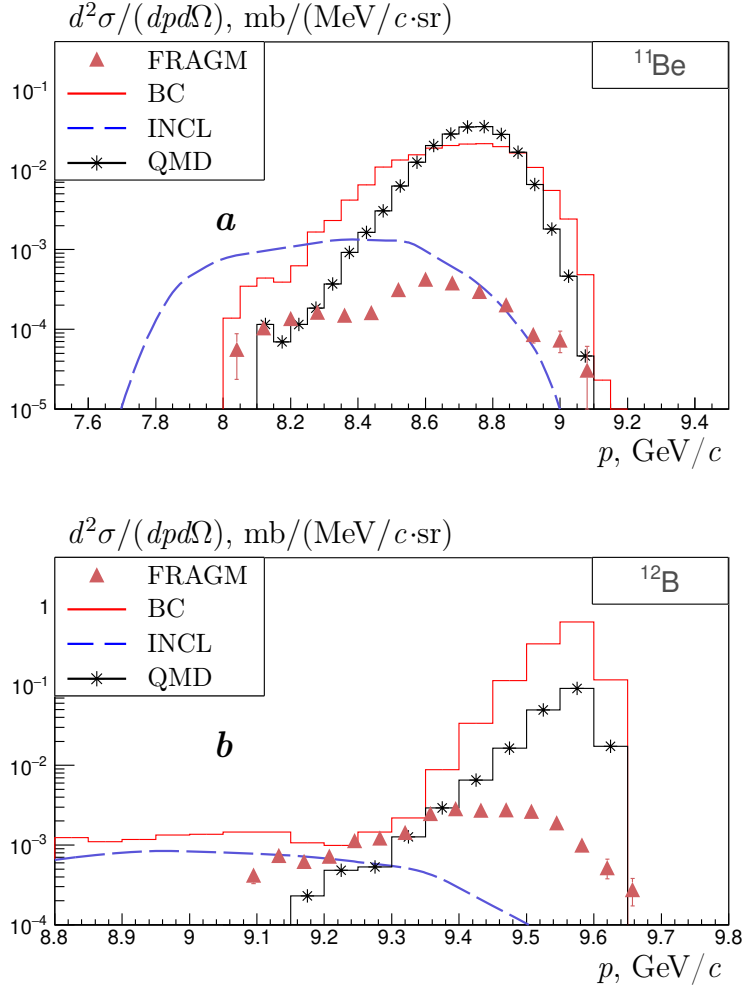
\begin{figure}[!htb]	
\includegraphics[scale=0.56]{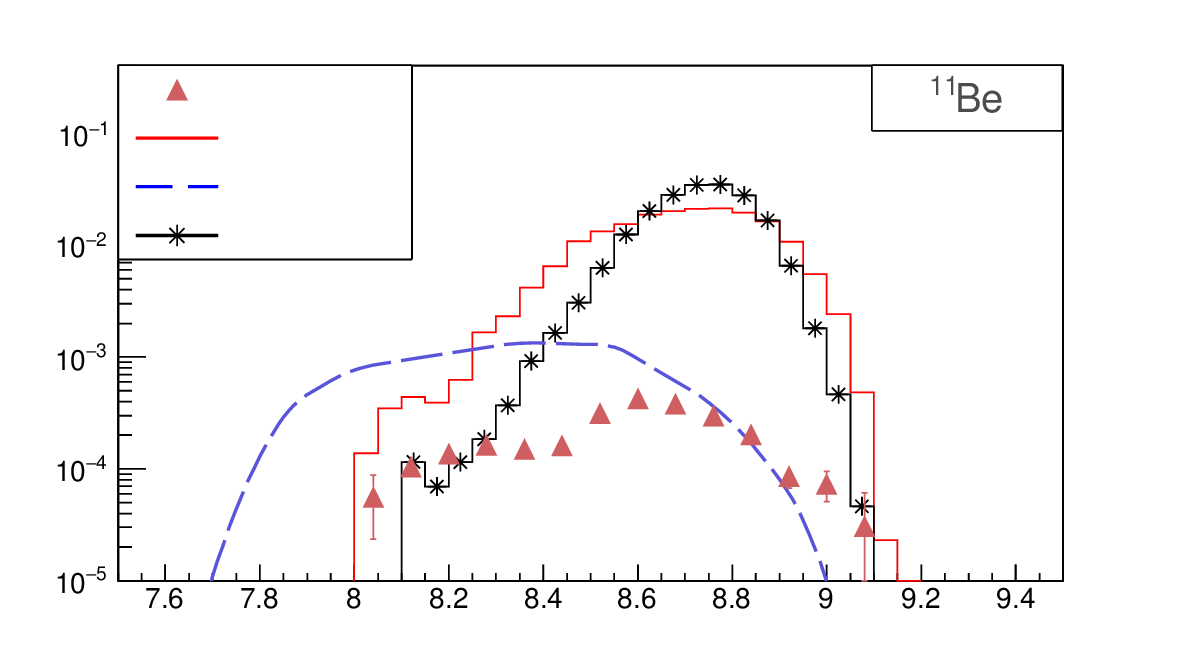}
\setlength{\unitlength}{1.5mm} 
\begin{picture}(400,0)	
\put(26,44){\fontsize{12}{12}\selectfont{{\it d}$^{\hspace{0.1em}2}\sigma/$({\it dpd}$\Omega$), mb$/$(MeV$/${\it c}$\hspace{0.1em}\cdot$\hspace{-0.15cm} sr)}}
\put(73,5.5){\fontsize{12}{12}\selectfont{{\it p}, GeV/\it{c}}}
\put(33.2,40.3){\fontsize{10}{10}\selectfont{FRAGM}}
\put(33.2,37.4){\fontsize{10}{10}\selectfont{BC}}
\put(33.2,34.3){\fontsize{10}{10}\selectfont{INCL}}
\put(33.2,31.3){\fontsize{10}{10}\selectfont{QMD}}
\put(39,26){\fontsize{14}{14}\selectfont{\it \textbf{a}}}
\end{picture}\vspace{-0.2cm}
\includegraphics[scale=0.56]{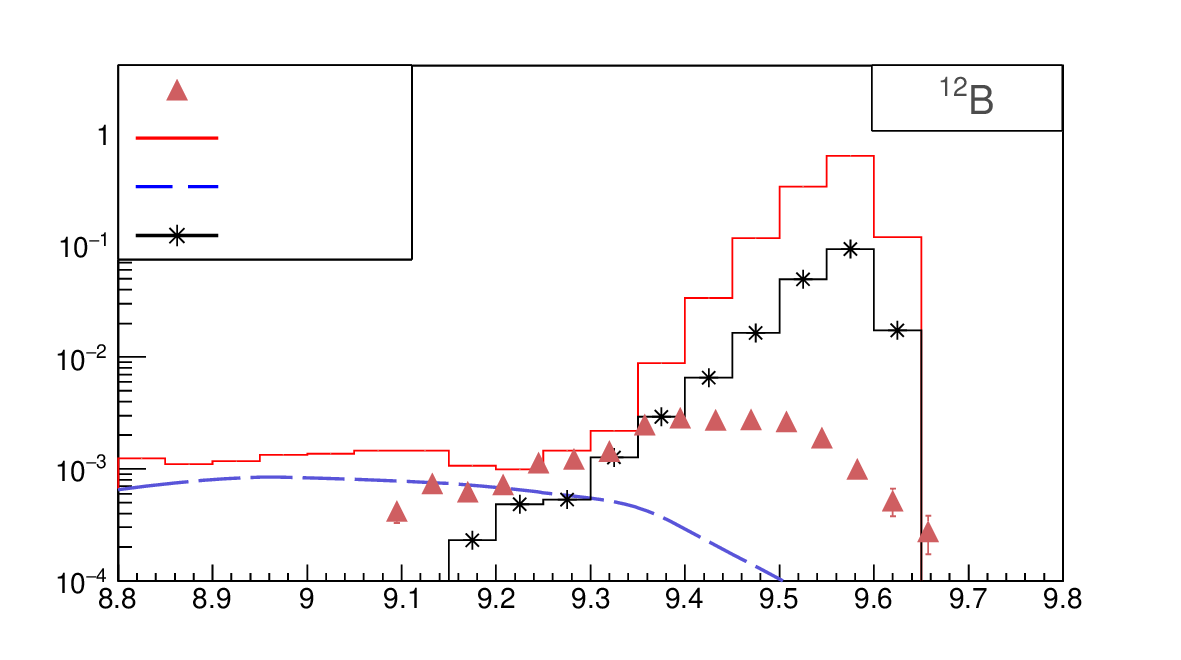}
\begin{picture}(400,0)
\put(26,44){\fontsize{12}{12}\selectfont{{\it d}$^{\hspace{0.1em}2}\sigma/$({\it dpd}$\Omega$), 
mb$/$(MeV$/${\it c}$\hspace{0.1em}\cdot$\hspace{-0.15cm} sr)}}
\put(73,5.5){\fontsize{12}{12}\selectfont{{\it p}, GeV/\it{c}}}
\put(33.2,40.3){\fontsize{10}{10}\selectfont{FRAGM}}
\put(33.2,37.4){\fontsize{10}{10}\selectfont{BC}}
\put(33.2,34.3){\fontsize{10}{10}\selectfont{INCL}}
\put(33.2,31.3){\fontsize{10}{10}\selectfont{QMD}}
\put(39,25){\fontsize{14}{14}\selectfont{\it \textbf{b}}}
\end{picture}\vspace{-0.5cm}
\caption{
Differential cross sections for the production of $^{11}$Be ({\it a}) and $^{12}$B ({\it b}) ions for experimental data and models of nucleus-nucleus interactions as functions of the fragment momentum.}\label{Pict10}
\end{figure}
Each channel setting gives an additional measurement corresponded to a certain hodoscope cell. Assuming the central cell corresponds to the rigidity of magneto-optical channel, it's possible to correlate the cell number with a certain rigidity value. Measurements at different channels adjustments and similar rigidity values are in a good agreement with each other, and that is a good prof of this method applicability. To determine the yields of fragments, the closest measurements in rigidity from different channel settings are summed up with a step of 10 -- 20 MeV/{\it c}.

Fig.~\ref{Pict10} shows the differential cross sections $d^2\sigma$/($d\Omega dp$) for the production of $^{11}$Be and $^{12}$B fragments as a function of the laboratory momentum. Experimental data are compared with similar spectra obtained from the BC, INCL, and QMD nucleus-nucleus interaction models. The obtained cross sections are smaller by more than two orders of magnitude than the cross sections for the yield of other isotopes produced without charge-exchange. 
The experimental shapes of the $^{12}$B and $^{11}$Be distributions have a Gaussian shape with a narrower width, 100 -- 150~MeV/{\it c}, compared to the shapes for boron and beryllium ions producing without charge exchange. The BC and QMD models are in satisfactorily agreement with the results of FRAGM experiment in terms of the mean value and width of the distribution, but they diverge from the experimental data in absolute values. The INCL model does not describe the characteristic peak corresponding to a single nucleon charge-exchange and predicts only the low-energy part of the momentum distribution.

\subsection{Calculation of the upper limit of $^{12}$N production. }

The production of isotopes $^{12}$N and $^{12}$B during the fragmentation of $^{12}$C, occurs due to the dominant process of one-pion exchange at intermediate energies. Theoretically, their cross-sections should be equal due to the fact that both fragments are produced on an isotope-symmetric target. However, under the experimental conditions with a beryllium target, which has less protons than neutrons, the probability of positive charge-exchange decreases
suppressing the formation of $^{12}$N in comparison to $^{12}$B. It is hardly possible to estimate this effect precisely, but under the simplest assumption that all nucleons of the target participate in the fragmentation reaction with equal probability, the coefficient of this suppression is equal to the ratio of the number neutrons to the number of protons in the target. This value is only 1.25 for a beryllium target. In addition, the experiment registers only long-lived isotopes that flying a significant distance. For the ground states of $^{12}$B and $^{12}$N, this is done with a large reserve, since their half-life time are 20.2 and 11.0~ms~\cite{PaperLT}.

In ion-ion interactions, fragments are produced not only in the ground state, but also in the excited state. 
Generally, transitions from excited states to the ground one occur due to electromagnetic interaction 
with the emission of soft gamma rays within a time around 10$^{-16}$~s, which allows to ignore the role of excited states. For $^{12}$B and $^{12}$N, the situation is more complicated. In the case of $^{12}$B, the excited states with an energy greater than 3.37~MeV are decaying due to the strong interaction within 10$^{-20}$~s via the $^{11}$B~+~$n$ channel, and the isotope is not registered by the setup in this case. Under an excitation energy less than 3.37~MeV, there are six levels, including the ground, which contribute to the producing of long-lived ground state of $^{12}$B. The threshold excitation energy of $^{12}$N for the decay to $^{11}$C~+~$p$ is only 0.601~MeV, which is less than the energy of first excited state of $^{12}$N(2$^+$) equal to 0.961~MeV. This means that all excited states of $^{12}$N decay to $^{11}$C and are not detected by the setup. Spectrum of different states of $^{12}$N was obtained in the nucleon charge-exchange reaction for the $^{12}$C($p$, $n$)$^{12}$N on a resting carbon nucleus. The measurements were carried out on the Indiana University cyclotron at proton energy of 135~MeV \cite{PaperReview8}. Neutrons were 
detected using scintillation detectors at angles of 0$^\circ$, 24$^\circ$ and 45$^\circ$ relative to the direction of initial proton beam. Based on these data, the spectrum of $^{12}$N nuclear levels was determined for each angle. 
The production of ground state dominates at an angle of 0.2$^\circ$, and for angle equal to 24$^\circ$ the contribution
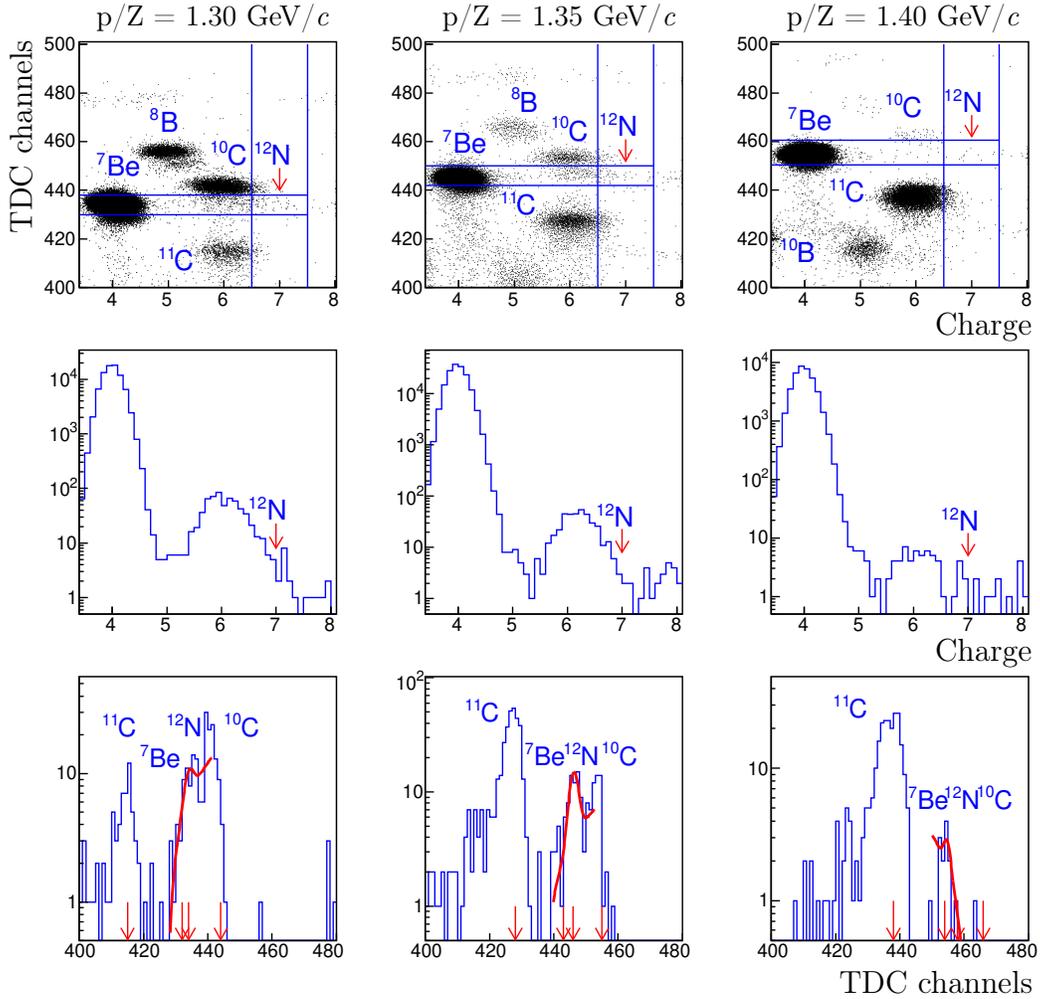
\begin{figure}[!htb]
\includegraphics[scale=0.69]{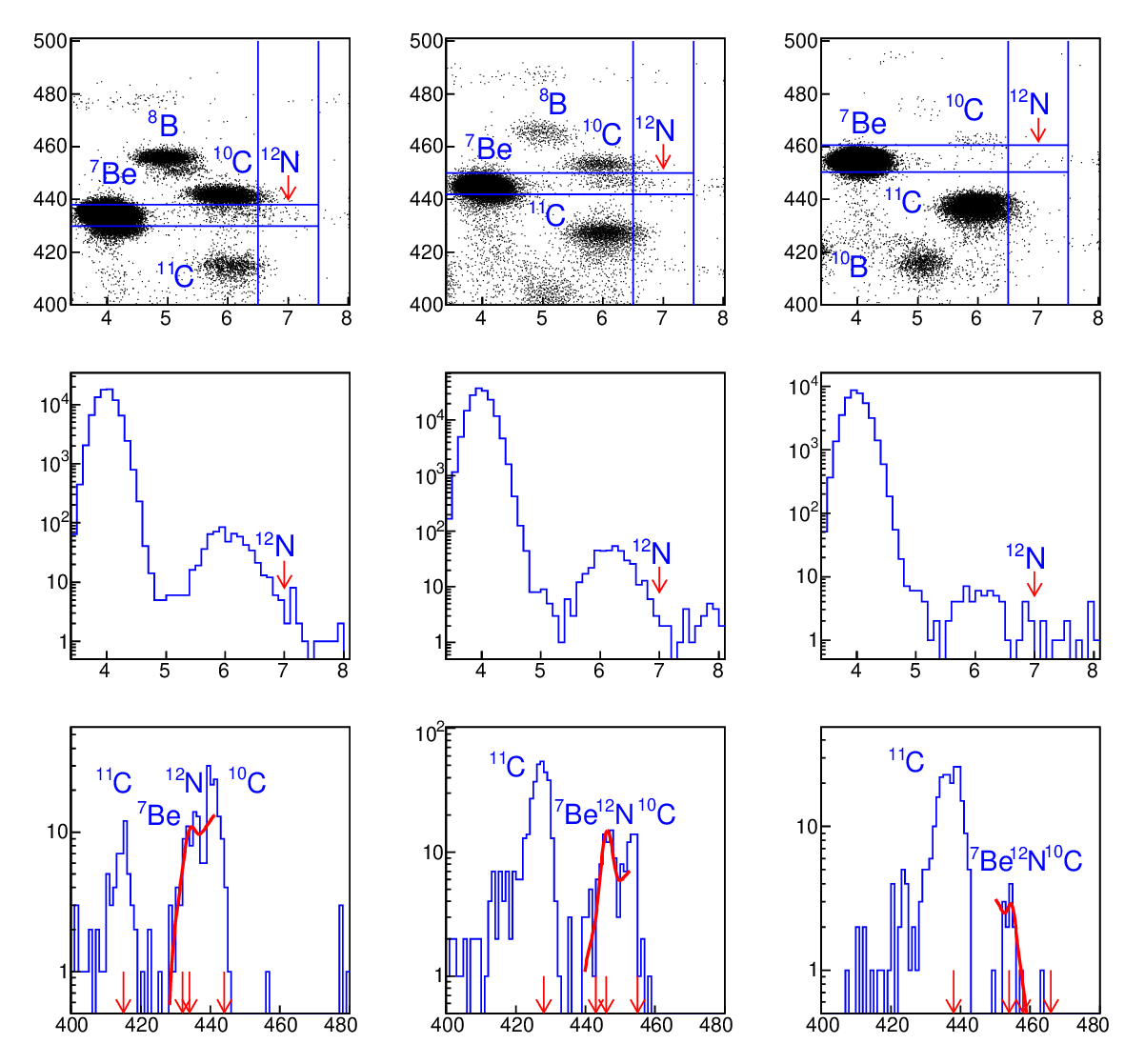}
\setlength{\unitlength}{1.5mm} 
\begin{picture}(400,0)
\put(14.8,90.4){\fontsize{11}{11}\selectfont\hspace{0.2cm}{p/Z = 1.30 GeV/{\it c}}}
\put(45.5,90.4){\fontsize{11}{11}\selectfont\hspace{0.2cm}{p/Z = 1.35 GeV/{\it c}}}
\put(76.0,90.4){\fontsize{11}{11}\selectfont\hspace{0.2cm}{p/Z = 1.40 GeV/{\it c}}}
\put(8.5,72){\fontsize{12}{12}\selectfont\rotatebox{90}{TDC channels}}
\put(90.5,63.0){\fontsize{12}{12}\selectfont{Charge}}
\put(90.5,34.0){\fontsize{12}{12}\selectfont{Charge}}
\put(82.0,4.6){\fontsize{12}{12}\selectfont{TDC channels}}
\end{picture}\vspace{-0.5cm}	
\caption{Correlation distributions for TDC and charge (upper row), charge distributions (middle row) and 
TDC distributions (lower row).}\label{Pict11}		
\end{figure}
of the ground state will be much smaller than the excited ones. In order to perform a comparison with our measurements, this reaction should be analyzed in inverse kinematics, in which the target nucleus is a proton and the projectile nucleus is carbon. In this reference frame, the emission angle of $^{12}$N will be approximately 12 times smaller than the given values. In our experiment, due to the larger angle and higher energy of the projectile ion, one can expect an even smaller contribution of the $^{12}$N ground state to the total yield of this isotope. Currently, it is not possible to estimate numerically this effect.

The relative probability of population of the $^{12}$N ground level to population of all levels could be determined from the data of FRAGM experiment, based on the equality of the $^{12}$B and $^{12}$N yields, should it be possible to register a signal from $^{12}$N. However, the background conditions for these measurements are unfavorable, and it was only possible to determine the upper limit of the cross section value for this isotope. Fig.~\ref{Pict11} shows the correlation distributions of the time-of-flight and charge for different adjustments of the magneto-optical channel by rigidity (upper row), as well as the distributions by charge (middle row) and by time-of-flight (lower row). The selection of $^{12}$N isotopes was carried out in the region of rigidity from 1.25 to 1.50~GeV/$\it{c}$, which ensures the maximum isotope yield. The amplitude, recalculated into the fragment charge, allows to align the signals from fragments with the same charge and to simplify the ion selection procedure. The events, which are inside the region of charge number equal to 7, were selected, and by the time-of-flight variable -- in the range of $\pm$ 5 channels (1~ns) from the calculated time-of-flight at a selected rigidity. The maximum in the region which corresponds to the time-of-flight $^{12}$N is well separated from the background peaks associated with carbon isotopes, but it is impossible to distinguish it from the background caused by $^7$Be. Therefore, all events forming the above-mentioned maximum, must be taken as a value of the upper limit of the nucleus yield. 

\begin{figure}[!htb]	
\includegraphics[scale=0.54]{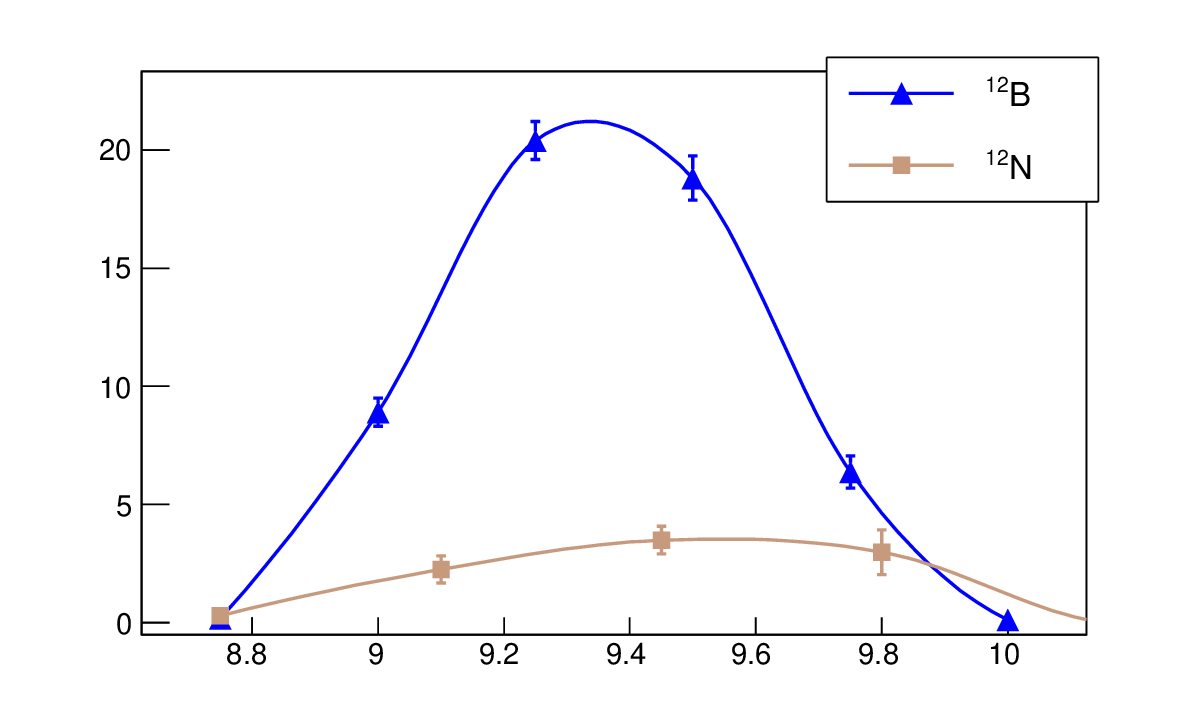}
\setlength{\unitlength}{1.5mm} 
\begin{picture}(400,0)
\put(70.5,5.5){\fontsize{12}{12}\selectfont\hspace{0.5cm}{{\it p}, GeV/{\it c}}}
\put(29.5,45.8){\fontsize{12}{12}\selectfont{Relative isotope yield}}
\end{picture}\vspace{-0.8cm}
\caption{Comparative ratio of the yields of isotope $^{12}$B and possible yield of $^{12}$N ground state for the upper limit estimating of detection $^{12}$N by the experimental setup.}\label{Pict12}
\end{figure}
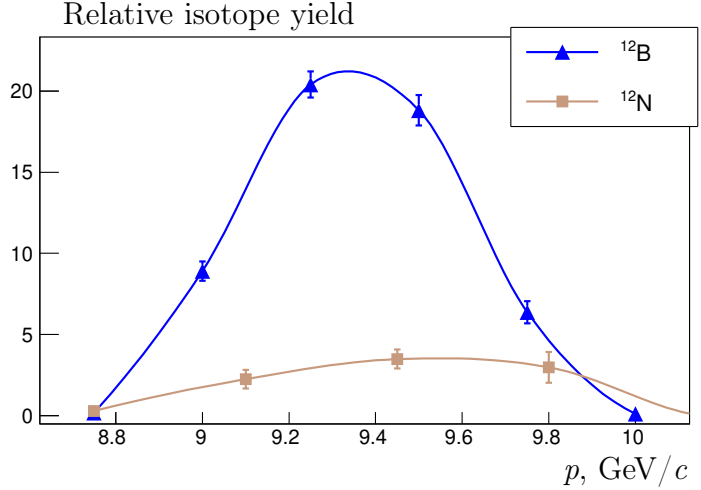

The relative yields of the isotopes $^{12}$B and $^{12}$N as functions of the ion momentum are shown in Fig.~\ref{Pict12}. The presented distributions use the detecting efficiency of the isotopes $^{12}$B and $^{12}$N. The suppression coefficient of $^{12}$N with respect to $^{12}$B was calculated at the momentum corresponding to the maximum yield of $^{12}$B and was about 7, indicating a significant suppression of the ground state of $^{12}$N in the reaction $^{12}$C(p, n)$^{12}$N measured at an angle of 3.5$^\circ$.

\subsection{Search for fragments obtained as a result of double nucleon charge-exchange. }

\begin{figure}[!htb]
\includegraphics[scale=0.56]{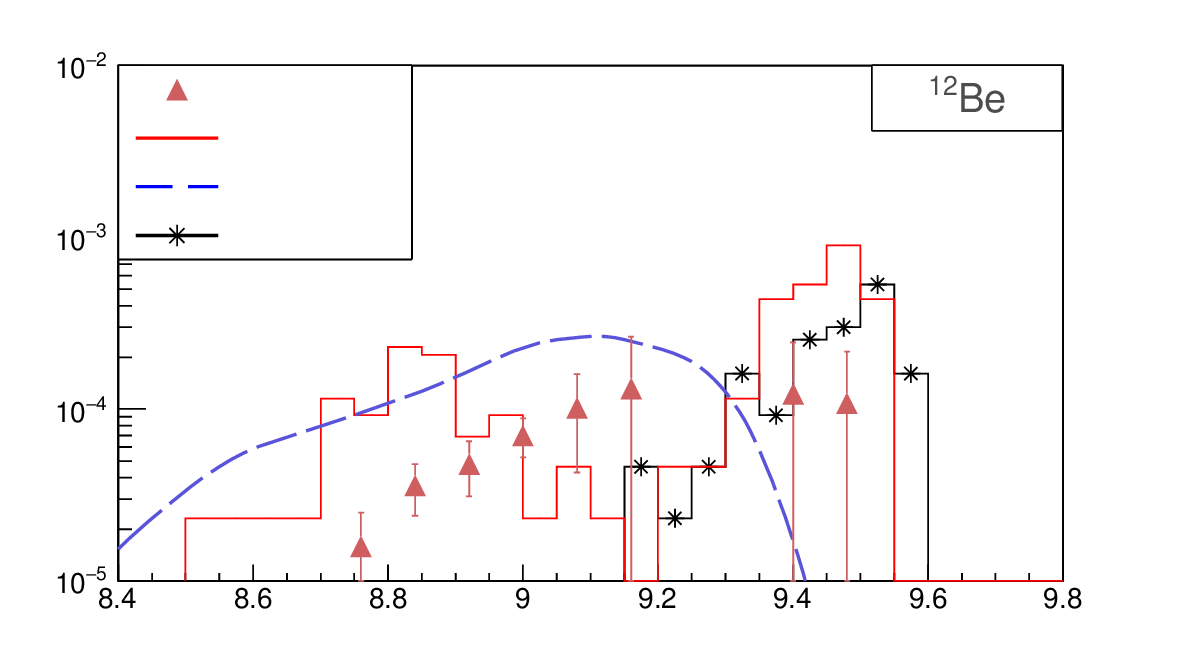}
\setlength{\unitlength}{1.5mm}
\begin{picture}(400,0)
\put(26,44){\fontsize{12}{12}\selectfont{{\it d}$^{\hspace{0.1em}2}\sigma/$({\it dpd}$\Omega$), mb$/$(MeV$/${\it с}$\hspace{0.1em}\cdot$\hspace{-0.15cm} sr)}}
\put(73,5.5){\fontsize{12}{12}\selectfont{{\it p}, GeV/\it{c}}}
\put(33.2,40.3){\fontsize{10}{10}\selectfont{FRAGM}}
\put(33.2,37.4){\fontsize{10}{10}\selectfont{BC}}
\put(33.2,34.3){\fontsize{10}{10}\selectfont{INCL}}
\put(33.2,31.3){\fontsize{10}{10}\selectfont{QMD}}
\end{picture}\vspace{-0.8cm}
\caption{Differential cross sections of $^{12}$Be for experimental data and predictions of the nucleus-nucleus interactions models as a function of the fragment momentum.}\label{Pict13}
\end{figure}
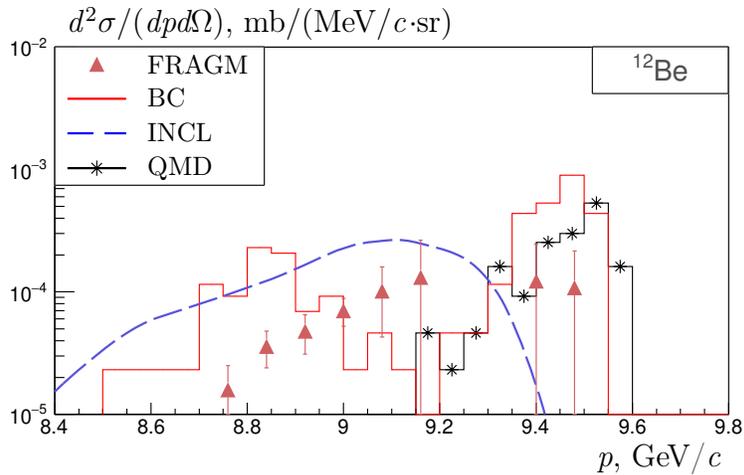

During the search of double nucleon charge-exchange, the $^{12}$Be isotope was discovered in the process 
which is accompanied a decrease of charge by two units while maintaining the mass number. Correlation dependencies of the time-of-flight and charge for various adjustments of the magneto-optical channel and the distributions of selected events over the hodoscope cells were analyzed in the rigidity range from 2.15 to 2.45~GeV/$\it{c}$. The cross section measured by us is presented in Fig.~\ref{Pict13} at comparison with the predictions of three ion-ion interaction models. A splitting of the momentum peak into two components was found.

The first peak, located in the region of momentum $p\sim$ 9.5~GeV/$\it{c}$, corresponds to the contribution from exchange of two virtual $\pi$-mesons. It is matching to the rigidity of 2.40~GeV/$\it{c}$, where only a few such events were detected, so its statistical significance is too low. On the other hand, it is clear that these experimental points 
are in a good agreement with the predictions of BC and QMD models. The next peak, located in the region of lower momentum values, has more statistics. It corresponds to the contribution of one virtual and real pion, which is  confirmed by the BC and INCL models. The BC model describes the picture of such a division most accurately, and the contributions from both peaks are approximately comparable to each other~\cite{PaperFRAGM2}.

\subsection{Statistical models of fragmentation processes. }

Statistical models of fragmentation are often used to estimate the shape of measured momentum spectra. Within the framework of these models, the momentum spectra in the rest frame of projectile nucleus are described by a Gaussian distribution. In the statistical model of Goldhaber~\cite{PaperTheor1,PaperTheor2}, the dispersion of momentum distribution measured at zero angle is determined 
through the mass numbers of the projectile nucleus $A_p$ and the fragment $A_f$ by the formula:
\begin{equation}\vspace{0.2cm}
\sigma_{||}^2=\sigma_0^2\frac{A_f(A_p-A_f)}{A_p-1} \\
\label{equation1}
\end{equation}
The constant $\sigma_0$ is expressed by the Fermi momentum of nucleons in nucleus $p_F$ as $\sigma_0 = p_F/\sqrt{5}$. Formula~(\ref{equation1}) predicts a parabolic dependence of $\sigma_{||}$ on the fragment mass number with a maximum at $A_f$ = $A_p$/2.

It should be noted, that the Goldhaber model was repeatedly modified and refined. Its main provisions that the motion of nucleons in the nucleus is not correlated and that nucleons in the fragment are chosen randomly were revised in different works. In the work~\cite{PaperTheor3} was shown the influence of Pauli principle on the momenta of nucleons consisting fragment. According to this principle, when two identical fermions are close to each other, they should have large anticorrelations in momenta. In particular, it was shown that during $^{40}$Ar fragmentation, the dispersion of momentum spectra is at least 30\% less than predicted by the conventional statistical model~\cite{PaperTheor4}. 
In~\cite{PaperTheor5}, it was demonstrated that the nucleons in projectile nucleus cannot be chosen completely randomly, provided that the nucleons in final fragment are determined by the Fermi gas model. It limits the available configurations of the nucleus and, therefore, reduces the predicted width of the momentum distribution. In the framework of peripheral model proposed in~\cite{PaperTheor6}, it was suggested to consider the projectile nucleus as an object consisting of two nucleon clusters: the fragment $F$ and the remainder $R$. Momentum dispersion in the rest frame of projectile nucleus is determined by the formula:
\begin{equation} 
\sigma_{||}^2 = \frac{\mu}{2x_0}\left[\frac{1+y/2}{\sqrt{1+y}} + \frac{1}{\mu x_0}\right],
\label{equation2}
\end{equation}
where $\mu = \sqrt{2m_R E_S}$ -- the reduced mass of the fragment and remainder, $E_S$ -- the energy of nucleon separation in the nucleus, $y~=~Z_1Z_2e^2/x_0Es$, $x_0~=~r_0A_f^{1/3}$. Thus, the momentum distribution of the fragment directly depends on the energy $E_S$, and not on the Fermi momentum energy. In general, it was shown that the introduction of parameter $E_S$ improves the agreement of experimental data with the model predictions. The predictions of this theory are similar to a simple statistical model, i.e. they indicate a parabolic dependence of the dispersion of the momentum distribution on the fragment mass.

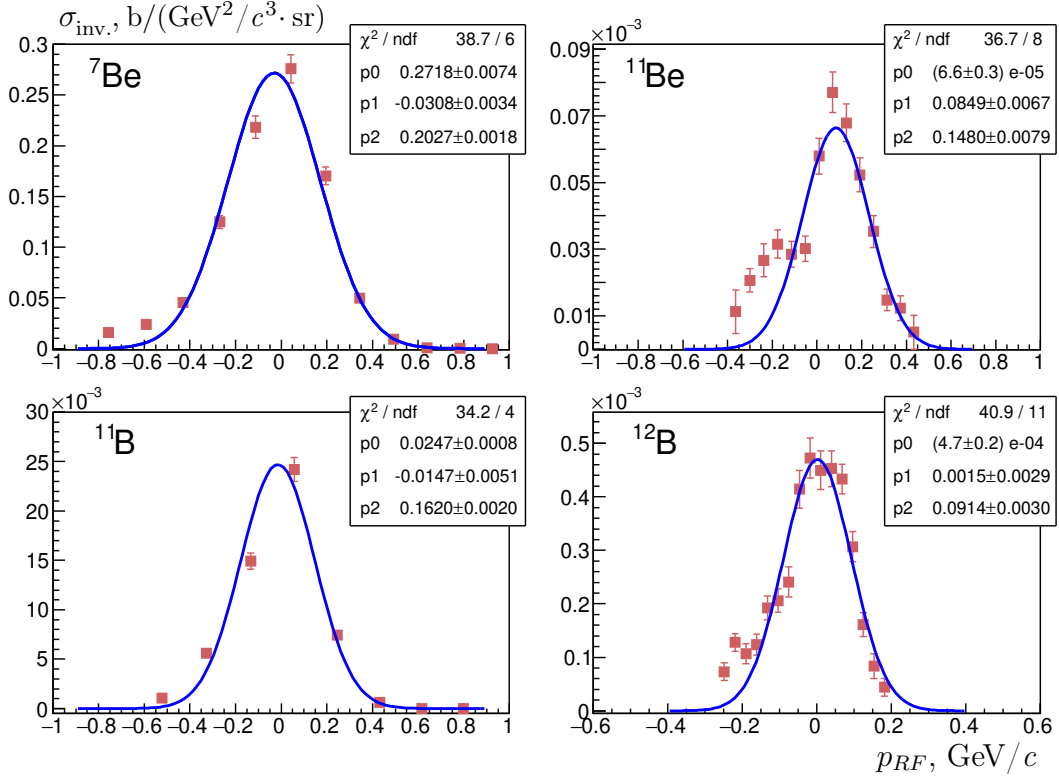
\begin{figure}[!htb]
\includegraphics[scale=0.72]{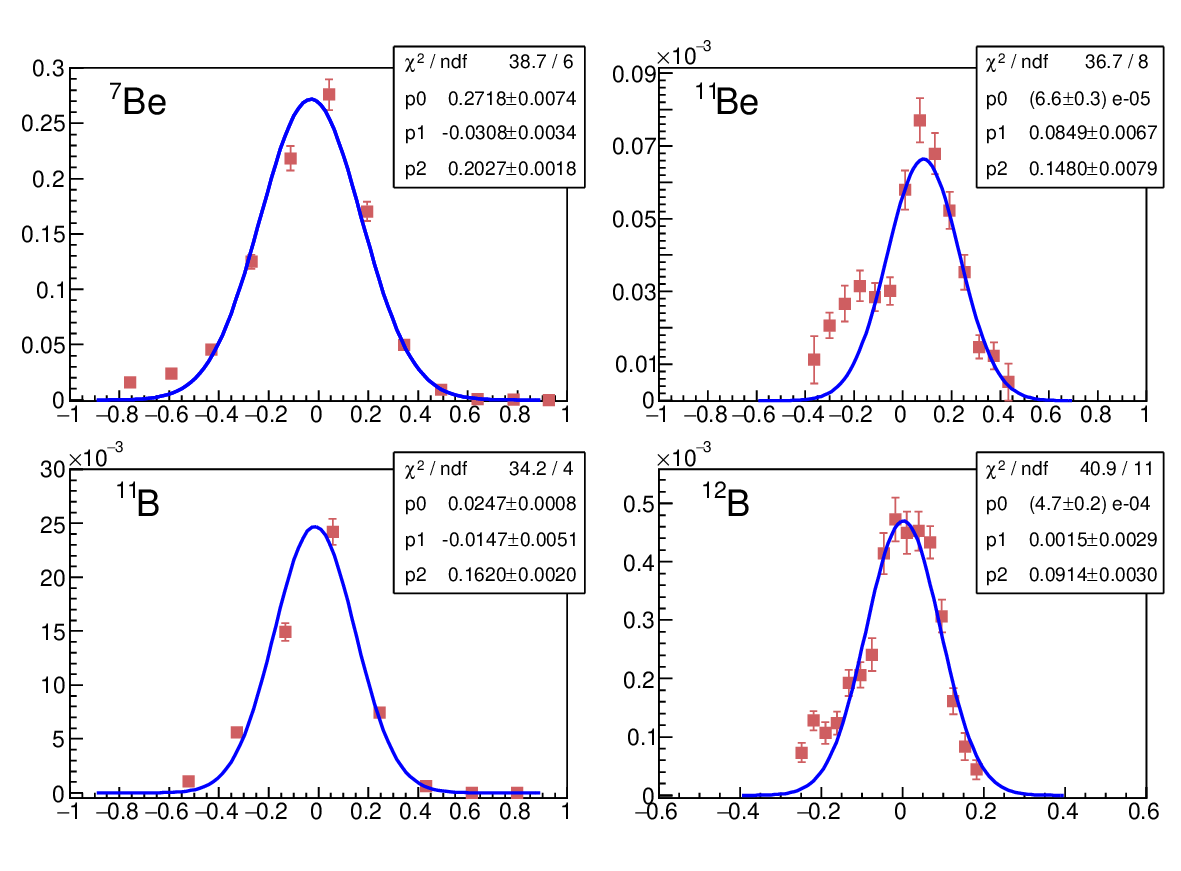}
\setlength{\unitlength}{1.5mm}
\begin{picture}(400,0)
\put(13,72.0){\fontsize{12}{12}\selectfont{$\sigma$\hspace{0.45cm},}}
\put(19,72.0){\fontsize{11}{11}\selectfont{b$/$(GeV$^{2}$/{\it c}$^{\hspace{0.1em}3}\cdot$\hspace{-0.05cm} sr)}}
\put(14.5,71.2){\fontsize{8}{8}\selectfont{inv.}}
\put(85.5,6.4){\fontsize{12}{12}\selectfont{$p_{RF}$, GeV/\it{c}}}
\end{picture}\vspace{-1.0cm}
\caption{Distribution of the invariant cross section for the production of beryllium and boron isotopes depending on 
the fragment momentum in the rest frame of incident nucleus.} \label{Pict14}
\end{figure}

The experimental distributions of $^7$Be, $^{11}$Be, $^{11}$B and $^{12}$B yields as a function of the $p_{RF}$ momentum calculated in the rest frame of the projectile carbon nucleus are shown in Fig.~\ref{Pict14}. The yield of fragments is represented by the invariant cross section:
\begin{equation}
\sigma_{inv.} = \frac{E}{p^2}\frac{d^2\sigma}{d\Omega dp}
\label{equation3}
\end{equation}

The measured widths of the momentum spectra $\sigma^{\text{\fontsize{6}{6}\selectfont{FRAGM}}}$ are determined by fitting the distribution data using a Gaussian function.

Since the width in formula~(\ref{equation1}) is calculated at zero angle, two corrections must be introduced to compare the experimental data with the model predictions. First correction is related to the momentum resolution of the magneto-optical path, which is 1.5\% of the momentum in laboratory frame. When it is important to achieve high accuracy to describe the spectra, information obtained from the hodoscope should be used. In this case, the correction is small and equal to 0.4\% of the fragment momentum. Second correction is associated with transformation the experimental width to zero angle. This correction was made using the BC model. These corrections were allowed to obtain the corresponding widths $\sigma_{||}^{\text{\fontsize{6}{6}\selectfont{FRAGM}}}$ for beryllium and boron isotopes.

The experimental data are shown in Table~\ref{TableOne} in comparison with similar data ($\sigma_{||}^{\text{exp}}$) given in the work where fragmentation of carbon ions with an energy of 2.1~GeV/nucleon on a beryllium target was studied~\cite{PaperTheor7}.
\begin{table}[!htb]\begin{center}\vspace{0.2cm}
\begin{tabular}{ 
|>{\hspace{0.1cm}\raggedright}p{1.5cm}|>{\centering\arraybackslash}p{2cm}|
 >{\centering\arraybackslash}p{2cm}|>{\centering\arraybackslash}p{2cm}|
 >{\centering\arraybackslash}p{2cm}|>{\centering\arraybackslash}p{2cm}|}
\hline
Isotope & $\sigma^{\text{\fontsize{6}{6}\selectfont{FRAGM}}}$,& $\sigma_{||}^{\text{\fontsize{6}{6}\selectfont{FRAGM}}}$,&   
$\sigma_{||}^{\text{exp}}$, &  $\sigma_{||}^{\text{theor1}},$ &  $\sigma_{||}^{\text{theor2}}$, \\
& МэВ/\it{с} & MeV/\it{c} & MeV/\it{c} & MeV/\it{с} & MeV/\it{c} \\  \hline
$^7$Be 	 & 203 $\pm$ 2 & 168 $\pm$ 6   & 145 $\pm$ 2   & 183  & 143 \\
$^9$Be   & 179 $\pm$ 2 & 137 $\pm$ 8   & 133 $\pm$ 3   & 161   & 128 \\
$^{10}$Be& 208 $\pm$ 1 & 131 $\pm$ 6   & 129 $\pm$ 4   & 138   & 124 \\
$^{11}$Be& 148 $\pm$ 8 & 111 $\pm$ 13  & 155 $\pm$ 4   & 103   & 123 \\
$^8$B    & 213 $\pm$ 3 & 180 $\pm$ 6   & 151 $\pm$ 16  & 175   & 154 \\
$^{10}$B & 206 $\pm$ 4 & 133 $\pm$ 6   & 134 $\pm$ 3   & 138   & 120 \\
$^{11}$B & 162 $\pm$ 2 & 100 $\pm$ 7   & 106 $\pm$ 4   & 103   & 102 \\
$^{12}$B & 91 $\pm$ 3  & 55 $\pm$ 4    & 64 $\pm$ 9    & --    & --  \\   
\hline
\end{tabular}\vspace{0.3cm}
\caption{Momentum standard deviation $\sigma$ for beryllium and boron isotopes~\cite{PaperFRAGM1}.}
\end{center}\label{TableOne}\end{table}
In addition, the widths obtained using Goldhaber's formula ($\sigma_{||}^{\text{theor1}}$) and calculated within the peripheral model ($\sigma_{||}^{\text{theor2}}$ by the formula~(\ref{equation2}) are given. The experimental values of $\sigma_{||}^{\text{\fontsize{6}{6}\selectfont{FRAGM}}}$ are in agreement with the widths obtained using the theoretical calculations. In the case of $^{12}$B, the heoretical predictions are not applicable, since they require the presence of a fragment and corresponding fraction from the projectile ion.

\section{CONCLUSION}

This work is devoted to analyzing the results of experimental measurements of carbon ion fragmentation processes occurring with nucleon charge-exchange at an energy of 300~MeV/nucleon obtained on the FRAGM detector and the TWAC accelerator. During the experimental data processing, the selection of investigated fragments was carried out according to the correlation distributions of the time-of-flight and amplitude at different settings of the magneto-optical channel for rigidity. A method was developed for improving the accuracy of measured momentum using information coming from the hodoscope system of the FRAGM detector. This procedure allowed to improve the accuracy of momentum measurement to 0.4\%, which is necessary to obtain the exact shape of the momentum spectra. Simulation the ion beam transportation in the magneto-optical channel of the FRAGM detector allowed to determine the main physical parameters of the detector, and also the fragment detection efficiency as a function of its momentum.

The isotopes $^{11}$Be and $^{12}$B were discovered producing as a result of single nucleon charge-exchange. 
The yield of isotope $^{12}$Be, produced in the process of double charge-exchange, was also measured. Differential cross sections for the isotopes production were obtained depending on their momentum. These cross sections values are 
two orders of magnitude smaller than the cross sections of isotopes produced without charge-exchange. Despite the yield of $^{12}$N could not be measured, an estimation of the upper limit of the yield of $^{12}$N was done relative to the yield of $^{12}$B.

The experimental data were compared with three models of nucleus-nucleus interactions: BC, INCL, QMD.
The comparison of obtained data with the model predictions data showed a difference in the description of momentum spectra shape for different models, including a difference with the experiment. In particular, the INCL model does not reproduce the quasielastic peak of charge-exchange reactions, the BC model reproduces the experimental data in shape by best possible way, but overestimates the yields of fragments. The shape of beryllium and boron momentum spectra width was estimated in the rest frame of the projectile nucleus. Experimental widths were compared with the predictions obtained within the Goldhaber and peripheral models, as well as with similar data on $^{12}$C fragmentation. In general, the experimental data are in good agreement with other data on fragmentation within the error limits. The new experimental data in present study allows extending information on nuclear reactions and the structure of light nuclei, which is important for model calculations and understanding the mechanisms of nuclear interactions.

\section*{Acknowledgements}
The authors are grateful first of all to V.~V.~Kulikov and P.~I.~Zarubin for their time in discussing the contents of article and for valuable comments, as well as to the staff of TWAC accelerator complex and the technical staff of the FRAGM experiment for their contribution to the measurements.


\begin{thebibliography}{99}
		
\bibitem{PaperIntr1}
De~Napoli~M. {\it et~al.}~// Phys.~Med.~Biol. 2012. V.~{\bf 57}. P.~7651--7671. 
\bibitem{PaperIntr2} 
Lenske~H., Bellone~J.~I., Colonna~M., Lay~J.~A.~// Phys.~Rev.~C. 2018. V.~{\bf 98}. P.~044620. 
\bibitem{PaperIntr3} 
Frekers~D., Alanssari~M.~// Eur.~Phys.~J.~A. 2018. V.~{\bf 54}. P.~177. 	
\bibitem{PaperIntr4} 
Geissel~H. {\it et~al.}~// Nucl.~Instrum.~Meth.~B. 1992. V.~{\bf 70}. P.~286--297. 
\bibitem{PaperIntr5} 
Ichihara~T. {\it et~al.}~// Nucl.~Phys.~A. 1994. V.~{\bf 569}. P.~287--296.
\bibitem{PaperIntr6}
Rodriguez-Sanchez~J.~L. {\it et~al.}~// Phys.~Lett.~B. 2020. V.~{\bf 807}. P.~135565.
\bibitem{PaperIntr7} 
Lenske~H., Wolter~H.~H., Bohlen~H.~G.~// Phys.~Rev.~Lett. 1989. V.~{\bf 62}. P.~1457--1460. 	
\bibitem{PaperIntr8} 
Amos~K., Karataglidis~S., Richter~W.~A.~// Eur.~Phys.~J.~A. 2020. V.~{\bf 56}. P.~284. 	
\bibitem{PaperIntr9} 
Kelic~A. {\it et~al.}~// Phys.~Rev.~C. 2004. V.~{\bf 70}. P.~064608.
\bibitem{PaperIntr10}
Abramov~B.~M. {\it et~al.}~// Phys.~At.~Nucl. 2022. V.~{\bf 85}. No.~9. P.~1541--1545.
\bibitem{PaperReview1}
Ellegaard~C. {\it et~al.}~// Phys.~Rev.~Lett. 1983. V.~{\bf 50}. P.~1745--1748.
\bibitem{PaperReview2}
Ableev~V.~G. {\it et~al.}~// JETP~Lett. 1984. V.~{\bf 40}. P.~763--766.
\bibitem{PaperReview3}
Contardo~D. {\it et~al.}~// Phys.~Lett.~B. 1986. V.~{\bf 168}. P.~331--335.
\bibitem{PaperReview4}
Skobelev~N.~K. {\it et~al.}~// Phys.~Part.~Nucl.~Lett. 2013. V.~{\bf 10}. P.~410--414.
\bibitem{PaperReview5}
Skobelev~N.~K. {\it et~al.}~// Phys.~Part.~Nucl.~Lett. 2014. V.~{\bf 11}. P.~114--120.
\bibitem{PaperReview6}
Skobelev~N.~K. {\it et~al.}~// Phys.~Part.~Nucl. 2022. V.~{\bf 53}. No.~2. P.~382--386.
\bibitem{PaperReview7}
Ichihara~T. {\it et~al.}~// Phys.~Lett.~B. 1994. V.~{\bf 323}. P.~278--283.
\bibitem{PaperReview8}
Anderson~B.~D. {\it et~al.}~// Phys.~Rev.~C. 1996. V.~{\bf 54}. P.~237--253.
\bibitem{PaperReview9}
Gareev~F.~A., Strokovsky~E.~A., Ratis~Yu.~L.~// Phys.~Part.~Nucl. 1994. V.~{\bf 25}. No.~9. P.~855--929.
\bibitem{PaperReview10} 
Rodriguez-Sanchez~J.~L. {\it et~al.}~// Phys.~Rev.~C. 2022. V.~{\bf 106}. P.~014618.
\bibitem{PaperReview11}
Mosbacher~C.~A., Osterfeld~F.~// Phys.~Rev.~C. 1997. V.~{\bf 56}. P.~2014--2028.
\bibitem{PaperReview12}
Tsuboyama~T., Sai~F., Katayama~N., Kishida~T., Yamamoto~S.~S.~// Phys.~Rev.~C. 2000. V.~{\bf 62}. P.~034001.
\bibitem{PaperSetup1}
Alekseev~N.~N., Koshkarev~D.~G., Sharkov~B.~Yu.~// JETP~Lett. 2003. V.~{\bf 77}. P.~123--125. 
\bibitem {PaperSetup2}
Alekseev~N.~N. {\it et~al.}~// Phys.~Part.~Nucl.~Lett. 2004. V.~{\bf 1}. P.~156--161.
\bibitem{PaperSetup3}
Satov~Yu.~A. {\it et~al.}~// Instrum.~Exp.~Tech. 2003. V.~{\bf 77}. P.~123--125. 
\bibitem{PaperSetup4}
Alekseev~N.~N. {\it et~al.}~// Atom.~Energy. 2003. V.~{\bf 95}. P.~794--800.
\bibitem{PaperSetup5}
Abramov~B.~M. {\it et~al.}~// JETP~Lett. 2013. V.~{\bf 97}. P.~439--443.
\bibitem{PaperSetup6} 
Abramov~B.~M. {\it et~al.}~// Phys.~Atom.~Nucl. 2015. V.~{\bf 78}. No~3. P.~373--380.
\bibitem {PaperSetup7}
Abramov~B.~M. {\it et~al.}~// J.~Phys.~Conf.~Ser. 2012. V.~{\bf 381}. P.~012037. 	
\bibitem {PaperSetup8}
Brun R., Rademakers~F.~// Nucl.~Instrum.~Meth.~A. 1997. V.~{\bf 389}. P.~81--86.
\bibitem {PaperSetup9}
Kharzheev Yu.~N.~// Phys.~Part.~Nucl. 2015. V.~{\bf 46}. No~4. P.~678--728. 
\bibitem {PaperSetup10}
Craun~R.~L., Smith~D.~L.~// Nucl.~Instrum.~Meth. 1970. V.~{\bf 80}. P.~239--244.
\bibitem {PaperSetup11}
Allison~J.~// Nucl.~Instrum.~Meth.~A. 2016. V.~{\bf 835}. P.~186--225.
\bibitem{PaperBC}
Folger.~G., Ivanchenko~V.~N., Wellisch~J.~P.~// Eur.~Phys.~J.~A. 2004. V.~{\bf 21}. P.~407--417.
\bibitem{PaperINCL}
Dudouet.~J., Cussol~D., Durand~D., Labalme~M.~// Phys.~Rev.~C. 2014. V.~{\bf 89}. P.~054616.
\bibitem{PaperQMD}
Aichelin~J.~// Phys.~Rept. 1991. V.~{\bf 202}. P.~233--360.
\bibitem{PaperFRAGM1} 
Kulikovskaya~А.~А. {\it et al.}~// Phys.~Atom.~Nucl. 2022. V.~{\bf 85}. No~5. P.~466--473.
\bibitem{PaperSihver}
Sihver~L. {\it et al.}~// Phys.~Rev.~C. 1993. V.~{\bf 47}. P.~1225--1236.
\bibitem{PaperLAQGSM}
Mashnik S.~G.~// Eur.~Phys.~J.~Plus. 2011. V.~{\bf 126}. P.~49.
\bibitem{PaperLT}
Kelley~J.~H., Purcell~J.E., Sheu~C.~G~// Nucl.~Phys.~A. 2017. V.~{\bf 968}. P.~71--253.
\bibitem{PaperFRAGM2} 
Kulikovskaya~А.~А. {\it et al.}~// Bull.~Russ.~Acad.~Sci.~Phys. 2023. V.~{\bf 87}. No~8. P.~1147--1150.
\bibitem{PaperTheor1}
Goldhaber A.~S.~// Phys.~Lett.~B. 1974. V.~{\bf 53}. P.~306--308. 
\bibitem{PaperTheor2}
Bacquias~A.~{\it et al.}~// Phys.~Rev.~C. 2012. V.~{\bf 85}. P.~024904.  
\bibitem{PaperTheor3}
Bertsch~G.~// Phys.~Rev.~Lett. 1981. V.~{\bf 46}. P.~472--473.
\bibitem{PaperTheor4}
Viyogi~Y.~P. {\it et~al.}~// Phys.~Rev.~Lett. 1979. V.~{\bf 42}. P.~33--36.
\bibitem{PaperTheor5}
Murphy M.~J.~// Phys.~Lett.~B. 1984. V.~{\bf 135}. P.~25--28.
\bibitem{PaperTheor6}
Friedman W.~A.~// Phys.~Rev.~C. 1983. V.~{\bf 27}. P.~569--577.
\bibitem{PaperTheor7}
Greiner D.~E.~// Phys.~Rev.~Lett. 1975. V.~{\bf 35}. P.~152.
		
\end{thebibliography}
\end{document}